\documentclass[twocolumn,amsmath,amssymb]{revtex4-1}

\usepackage{mathrsfs}
\usepackage{mathtools}
\usepackage{xfrac}
\usepackage{graphicx}
\usepackage{dcolumn}
\usepackage{bm}
\usepackage{tabularx}
\usepackage{amsmath}

\usepackage{dsfont}
\usepackage[mathscr]{eucal}
\usepackage[colorlinks, linkcolor=OrangeRed,citecolor=RoyalBlue,urlcolor=NavyBlue]{hyperref}
\usepackage[normalem]{ulem}
\usepackage[all]{hypcap}
\usepackage[dvipsnames,usenames,table]{xcolor}

\newcounter{defcounter}
\usepackage{amsthm}

\newcommand{\Tr}{\text{Tr}}
\newcommand{\bk}{{\bf k}}
\newcommand{\bq}{{\bf q}}
\newcommand{\bK}{{\bf K}}

\newcommand{\bd}{{\bf d}}

\newcommand{\bS}{{\bf S}}
\newcommand{\bp}{{\bf p}}
\newcommand{\br}{{\bf r}}
\newcommand{\vf}{v_F}
\newcommand{\kf}{k_F}
\newcommand{\ve}{{\varepsilon}}

\newcommand{\makebf}[1]{\boldsymbol{ #1 }}

\newcommand{\be}{\begin{equation}}
\newcommand{\ee}{\end{equation}}
\newcommand{\beq}{\begin{eqnarray}}
\newcommand{\eeq}{\end{eqnarray}}
\newcommand{\bea}{\begin{align}}
\newcommand{\eea}{\end{align}}
\newcommand{\beqq}{\begin{eqnarray*}}
\newcommand{\eeqq}{\end{eqnarray*}}

\newcommand{\up}{\uparrow}
\newcommand{\down}{\downarrow}

\newcommand{\ua}{\uparrow}
\newcommand{\da}{\downarrow}

\newcommand{\trs}{\Theta}
\newcommand{\phs}{\Xi}
\newcommand{\inv}{P}

\begin{document}

\begin{titlepage}

\title{Three-Dimensional Majorana Fermions in Chiral Superconductors}

\author{Vladyslav Kozii}
\affiliation{Department of Physics, Massachusetts Institute of Technology,
Cambridge, MA 02139, USA}

\author{J\"orn W. F. Venderbos}
\affiliation{Department of Physics, Massachusetts Institute of Technology,
Cambridge, MA 02139, USA}

\author{Liang Fu}
\affiliation{Department of Physics, Massachusetts Institute of Technology,
Cambridge, MA 02139, USA}

\begin{abstract}
Through a systematic symmetry and topology analysis  we establish that three-dimensional chiral superconductors with strong spin-orbit coupling and odd-parity pairing generically host low-energy nodal quasiparticles that are spin-non-degenerate and realize Majorana fermions in three dimensions. By examining all types of chiral Cooper pairs with total angular momentum $J$ formed by Bloch electrons with angular momentum $j$ in crystals, we obtain a comprehensive classification of gapless Majorana quasiparticles in terms of energy-momentum relation and location on the Fermi surface. 
We show that the existence of bulk Majorana fermions in the vicinity of spin-selective point nodes is rooted in the non-unitary nature of chiral pairing in spin-orbit-coupled superconductors. We address experimental signatures of Majorana fermions, and find that the nuclear magnetic resonance (NMR) spin relaxation rate is significantly suppressed for nuclear spins polarized along the nodal direction as a consequence of the spin-selective Majorana nature of nodal quasiparticles. Furthermore, Majorana nodes in the bulk have nontrivial topology and imply the presence of Majorana bound states on the surface that form arcs in momentum space. We conclude by proposing the heavy fermion superconductor PrOs$_4$Sb$_{12}$ and related materials as promising candidates for non-unitary chiral superconductors hosting three-dimensional Majorana fermions.
\end{abstract}


\maketitle

\draft

\vspace{2mm}

\end{titlepage}

\section*{Introduction \label{sec:intro}}

Chiral superconductors exhibit Cooper pairing with finite angular momentum thus spontaneously breaking time reversal symmetry \cite{kallin}. Two-dimensional chiral superconductors have been extensively studied in the context of Sr$_2$RuO$_4$ \cite{maeno}.
They are generally expected to have a  full superconducting gap and support topologically protected quasiparticles at the edge and in the vortex core.
In contrast, three-dimensional (3D) chiral superconductors generally have nodes. A well-known example is superfluid He-3 in the $p_x+ip_y$ paired $A$ phase, which has two point nodes on the Fermi surface along the $p_z$ axis \cite{leggett75}. Quasiparticles near these nodes are spin-degenerate and correspond to Weyl fermions \cite{balents,volovik-weyl}. Interestingly, when spin-orbit coupling is present, chiral superconductors with odd-parity (e.g., $p$-wave) pairing may have {\it non-unitary} gap structures and {\it spin-selective} point nodes \cite{venderbos15}. In this case, despite that the Fermi surface is spin degenerate, only states of one spin polarization at the nodal points are gapless in the superconducting state, whereas states of the opposite spin polarization are gapped.
Consequently, low-energy nodal quasiparticles arise from pairing within states of the same spin. These quasiparticles are identical to their antiparticles and thus are the solid-state realization of 3D Majorana fermions. 

Majorana fermions in condensed matter have recently attracted a great deal of attention \cite{wilczek,beenakker13}. Much of the studies so far has focused on localized Majorana fermion zero modes in quantum devices. In contrast, 3D Majorana fermions that are naturally occurring as itinerant quasiparticles in bulk chiral superconductors, have been little studied. In particular, it has been unclear what distinctive properties these Majorana quasiparticles have, and in what materials they are likely to be found.

In this work, 
we develop a systematic approach to classifying different types of Majorana
quasiparticles around spin-selective point nodes in chiral superconductors.
We present the criterion for such Majorana nodes on high-symmetry axis based on the symmetry of the superconducting order parameter and the band symmetry in the normal state. We further infer the presence of Majorana nodes away from high-symmetry axis from the topology of gap structures in the momentum space.
We show that the Majorana nature of nodal quasiparticles gives rise to a strongly anisotropic spin relaxation rate depending on the spin direction, which can be directly measured in NMR experiment.
Similar to Weyl fermions in topological semimetals, the presence of Majorana quasiparticles in chiral superconductors leads to a nodal topological superconductor phase which exhibits Majorana fermion surface states. As we demonstrate explicitly, zero-energy Majorana surface states form arcs in the surface Brillouin zone, which end at the bulk Majorana nodes. 
Finally we propose the heavy fermion superconductor PrOs$_4$Sb$_{12}$ as a promising candidate for chiral superconductor hosting Majorana quasiparticles.

\section*{Results}

\subsection*{Symmetry analysis of quasiparticle gap structures \label{sec:classification}}

We start with a general symmetry-based analysis of superconducting gap nodes in chiral superconductors with strong spin-orbit coupling and inversion symmetry. We assume that time reversal symmetry is present in the normal state and is spontaneously broken in the superconducting state due to the chiral pairing. We assume that chiral Cooper pairs carry a nonzero total angular momentum (including both orbital and spin) $J$ along a crystal axis  of $n$-fold rotation $C_n$, which acts jointly on electron's coordinate and spin. Here $n$ can only be $2,3,4, 6$ for discrete rotation symmetry of crystals and $J$ is only defined mod $n$. Moreover, since $\pm J$ corresponds to time-reversed chiral states, it suffices to consider positive integers $J=1,..., n/2$ for $n=2,4,6$, and $J=1$ for $n=3$.

In this work we address the gap structure associated with the points on the Fermi surface along the $n$-fold axis (hereafter denoted as $z$), whose momenta are given by $\pm \bK=\pm k_F \hat{z}$, where $k_F$ is Fermi momentum. Our approach to derive the gap structures relies on both symmetry and topological arguments. First, on the basis of a systematic symmetry analysis we show that the form of the gap structure at $\pm \bK$, i.e., the $C_n$-invariant Fermi surface momenta, is entirely determined by the total angular momentum $J$ of the Cooper pair and the angular momentum of energy bands at $\pm \bK$ in the normal state. Following the symmetry analysis, we invoke a topological constraint on the nodal structure of the quasiparticle spectrum to deduce the full low-energy gap structure, both at and away from $\pm \bK$. Using these two complementary methods we will demonstrate the existence of two types of point nodes, located on and off the $C_n$ axis respectively.

In the presence of both time-reversal ($\Theta$) and inversion ($P$) symmetries, spin-orbit-coupled energy bands remain two-fold degenerate at each momentum and we label the degenerate bands by a pseudospin spin index $\alpha=\up,\down$. For simplicity, we will simply refer to $\alpha$ as spin. The presence of $C_n$, $\Theta$ and $P$ symmetries guarantees that one can choose a basis for Bloch states at $\pm \bK$ such that {\it (i)} the state with
$\alpha=\up (\down)$ has angular momenta $j$ ($-j$), i.e.,
\beq
C_n c_{ \up(\down)} C_n^{-1}= e^{\pm i 2\pi j/n} c_{ \up (\down)},
\eeq
where $j$ is a positive half-integer; {\it (ii)} $P c_{\bK \alpha} P^{-1} = c_{-\bK \alpha}  $; {\it (iii)} $\Theta c_{\bK \alpha} \Theta^{-1} =\epsilon_{\alpha \beta} c_{-\bK \beta}  $, where $\epsilon_{\alpha \beta}$ is Levi-Civita symbol.

Having specified the angular momentum $J$ of the chiral Cooper pair and the angular momentum $\pm j$ of Bloch electrons, we are ready to deduce the gap structure near $\pm \bK$ by symmetry analysis. Only pseudospin-triplet pairings, which have odd-parity symmetry, may generate spin-dependent superconducting gaps necessary for 3D Majorana fermions.
There are three triplet pairing operators between states near $\pm \bK$, denoted by
\beq
\Gamma^{1}_\bq &=& c^\dagger_{\bK+\bq\up} c^\dagger_{-\bK-\bq\up}, \nonumber \\
\Gamma^2_\bq &=& c^\dagger_{\bK+\bq\down} c^\dagger_{-\bK-\bq\down}, \nonumber \\
\Gamma^3_\bq &=& ( c^\dagger_{\bK+\bq\up} c^\dagger_{-\bK-\bq\down} + c^\dagger_{\bK+\bq\down} c^\dagger_{-\bK-\bq\up}),
\eeq
where $\Gamma^{1,2,3}$ at $\bq=0$ carry angular momentum $2j$, $-2j$ and $0$, respectively.
In general the pairing potential near $\pm \bK$ is a mixture of these three pairing operators, with corresponding form factors
\beq
\mathcal H_p &=& \sum_\bq \sum_i \Delta_i (\bq) \Gamma^i_\bq + \text{H. c.}. \label{eq:delta_i}
\eeq
Since we are interested in the gap structure near $\bq=0$, it suffices to expand
 $\Delta_i(\bq)$ to the leading order in $\bq$:
\beq
\Delta_i (\bq) = C^+_i q_+^{a_i} + C^-_i q_-^{b_i},  \label{eq:deltaexpand}
\eeq
where we defined $q_{\pm} = q_x \pm i q_y$, and  $(q_x, q_y)$ is the momentum tangential to the Fermi surface at $\pm \bK$. The exponents $a_i, b_i$ are integers greater than or equal to zero. When $a_i \neq b_i$, the smaller of the two determines the leading order behavior of the gap function, while the other can be neglected. When $a_i = b_i$, both terms are equally important and should be kept together.

Importantly, the form of $\Delta_i(\bq)$ is constrained by the requirement that the pairing term $\mathcal H_p$ carries the angular momentum $J$. This completely determines the exponents $a_i, b_i$, i.e.,  the analytic form of $\Delta_i(\bq)$ at small $\bq$, allowing us to deduce the gap structures in the vicinity of $\pm \bK$.

Consider first the case $J=0$ mod $n$, i.e., when the superconducting order parameter has effectively zero angular momentum with respect to the $C_n$ rotation axis. In this case, the triplet pairing component with zero angular momentum $\Gamma_3$ is allowed at $\pm \bK$, i.e.,   $\Delta_3(\bq)$ is finite at $\bq=0$, creating a full pairing gap without any low-energy quasiparticles.

Next consider nonzero (mod $n$) $J$. If $J \neq 2j $ mod $n$, none of the three triplet pairing terms can be finite at $\pm \bK$, i.e., $\Delta_{i,\bq} \rightarrow 0$ as $\bq \rightarrow 0$ for all $i=1,2,3$. This implies that both spin $\up$ and $\down$ electrons are gapless at $\pm \bK$, resulting in spin-degenerate nodes at $\pm \bK$ and non-Majorana nodal quasiparticles. The low-energy Hamiltonian for such gapless quasiparticles can be determined from Eqs.~\eqref{eq:delta_i} and \eqref{eq:deltaexpand}.

\begin{table}[t]
\centering
\begin{ruledtabular}
\begin{tabular}{ccccc}
$C_n$  & $j$   & $J=2j$  & $l$ (mod $n$)  & Pairing $\Delta_\bq$ \\ [4pt]
\hline
$n=2$ & $j=\frac{1}{2}$ & $J=1$ & $l = 0$ &  $  \propto 1$  \\[5pt]
$n=3$ & $j=\frac{1}{2}$ & $J=1$ & $l = -1$ &  $  \propto q_-$  \\[5pt]
$n=4$ & $j=\frac{1}{2}$ & $J=1$ & $l = -2, 2$ &  $ \propto q^2_-,q^2_+ $  \\[5pt]
 & $j=\frac{3}{2}$ & $J=3$ & $l = -2, 2$ &  $ \propto q^2_-,q^2_+ $  \\[5pt]
 $n=6$ & $j=\frac{1}{2}$ & $J=1$ & $l = 2$ &  $  \propto q^2_+$ \\[5pt]
  & $j=\frac{3}{2}$ & $J=3$ & $l = 0$ &  $ \propto 1$ \\[5pt]
 & $j=\frac{5}{2}$ & $J=5$ & $l = -2$ &  $ \propto q^2_-$
\end{tabular}
\end{ruledtabular}
 \caption{{\bf Classification of pairing potentials.} Table summarizing the classification of pairing potentials $\Delta_\bq \equiv \Delta_{2,\bq}$ of the spin-$\down$ states $c^\dagger_{\pm\bK+\bq \down}$ to lowest order in $(q_+,q_-)$, with $q_\pm = q_x \pm i q_y$. The potentials are classified for given combination of $(n,j)$, where $n$ describes an $n$-fold rotation axis and $j$ is the spin angular momentum. The chiral superconductor has total angular momentum $2j$ and the effective orbital angular momentum of $\Delta_\bq$ is given by $l$.
 }
\label{tab:classification}
\end{table}


\textbf{\textit{Majorana nodes on rotation axis}}.  The type of chiral pairing giving rise to the Majorana nodal quasiparticles --- the focus of this work --- corresponds to $J=2j$ mod $n$. This implies odd $J$ and we exhaustively list all such cases in Table \ref{tab:classification}. Except for two cases $(n,j)=(2,\frac{1}{2})$ and $(6,\frac{3}{2})$ to be addressed separately later, we have $2j \neq -2j$ mod $n$. Under this condition, the spin $\up$ states carrying angular momentum $j$ are allowed to (and generally will) pair up  and form Cooper pairs carrying total angular momentum $2j$, while the spin $\down$ states remain gapless at $\pm \bK$ due to the angular momentum mismatch.
The resulting nodal quasiparticles are therefore spin non-degenerate Majorana fermions.

The low-energy Hamiltonian for these quasiparticles is given by
\begin{gather}
\mathcal H =\sum_{\bq} \xi_\bq ( c^\dagger_{\bq 1} c_{\bq 1} + c^\dagger_{-\bq 2} c_{-\bq 2} )
+  ( \Delta_\bq c^\dagger_{\bq 1} c^\dagger_{-\bq 2} + \text{H.c.} ), \label{eq:H1}
\end{gather}
where we have defined $c_{\bq 1,2} \equiv c_{\pm\bK+\bq \downarrow}$ and $\Delta_\bq \equiv \Delta_{2,\bq}$. In addition, $\xi_\bq \equiv \varepsilon_{\bK+\bq}-\mu$ where $\varepsilon_\bk$ is the single-particle energy-momentum relation and $\mu$ is the chemical potential. For small $\bq$ we have $\xi_\bq  = v_F q_z $, where $v_F = k_F/m$ is Fermi velocity in the $\hat z$ direction.

It is instructive to write $\mathcal H$ in Nambu space by introducing the four-component fermion operator $\Psi_\bq^\dagger$:
\beq
\Psi_\bq^\dagger = (c^\dagger_{\bq 1}, c^\dagger_{\bq 2}, c_{-\bq 1}, c_{-\bq 2} ), \label{eq:majoranafield}
\eeq
so that $\mathcal H$ can be expressed as
\beq
\mathcal H = \frac{1}{2} \sum_\bq \Psi_\bq^\dagger H(\bq) \Psi_\bq, \label{eq:Hlowenergy}
\eeq
with the $4\times 4$ matrix $H(\bq)$ taking the general form
\beq
H(\bq)= \begin{pmatrix}  \xi_\bq & 0 & 0 &  \Delta_\bq \\
0&  \xi_{-\bq}   &  -\Delta_{-\bq} &0\\
0 &  -\Delta^*_{-\bq}   &  -\xi_{-\bq}  & 0 \\
  \Delta^*_{\bq} &0 &0  &  -\xi_{\bq}
\end{pmatrix} . \label{eq:bdg}
\eeq
Importantly, the four-component quantum field $\Psi$ satisfies the same reality condition as Majorana fermions in high-energy physics, which reads as
$
\Psi^\dagger_\bq = (\tau_x \Psi_{-\bq})^T, \label{eq:reality}
$
in momentum space, or equivalently,
$
\Psi^\dagger_\br = ( \tau_x \Psi_\br )^T
$
in real space, where the Pauli matrix $\tau_x$ acts on Nambu space and $\Psi^T$ is the transpose of $\Psi$. This reality condition demonstrates that the low-energy quasiparticles can be regarded as Majorana fermions in three dimensions.

At small $\bq$, the pairing term $\Delta_\bq$ in (\ref{eq:bdg}) can be expanded in powers of $q_+$ or $q_-$. The exponent is determined by the mismatch between the angular momentum of the Cooper pair $J=2j$ and that of the spin $\down$ pairing operator $\Gamma_2$ at $\bq=0$ which is equal to $-2j$. Hence, one finds that
\beq
\Delta_\bq \propto [q_x + i \; \textrm{sgn}(l) q_y]^{|l|} \textrm{ with } l = 4j\mod n  \label{eq:delta_qform}
\eeq
The smallest allowed integer $|l|$ gives the form of $\Delta_\bq$ to the leading order. For any given $(n,j)$ and with $J=2j$ fixed, the corresponding $l$ is listed in Table \ref{tab:classification} (more details can be found in the Appendices).

From Table \ref{tab:classification}, we find three types of pairing terms $\Delta_\bq$ with different $l$'s, which give rise to two types of Majorana fermions with different energy-momentum relations. First, for $(n,j)=(3,\frac{1}{2})$ one has $l=1$ mod $n$, hence $|\Delta_\bq| \propto q_{\perp},$ where we defined $q_{\perp} = (q_x^2 + q_y^2)^{1/2}$. This implies that the quasiparticles near the nodes $\pm \bK$ disperse linearly with $\bq$ in all directions, as governed by the following effective Hamiltonian to first order in $\bq$,
\begin{gather}
H(\bq) =  v_F q_z \sigma_z + v_\Delta \sigma_x ( q_y\tau_x- q_x\tau_y ). \label{eq:linear}
\end{gather}
where  $\sigma_z =\pm 1$ denotes the two nodes $\pm \bK$, and $v_\Delta$ is defined via $|\Delta_\bq| = v_\Delta q_{\perp} + \mathcal{O}(q^2)$. Except for the velocity anisotropy, $H(\bq)$ is identical to the relativistic Hamiltonian for Majorana fermions in particle physics.

Second, we find several cases for which $l=\pm 2$ mod $n$. According to Eq. \eqref{eq:delta_qform},  this implies that the gapless quasiparticles disperse quadratically in $q_x, q_y$ and linearly in $q_z$ (see Table \ref{tab:classification}), as governed by the following effective Hamiltonian $H(\bq)$ to second order in $\bq$:
\begin{gather}
H(\bq) = v _F q_z \sigma_z + \frac{1}{2m_\Delta}\sigma_y[(q^2_x-q^2_y)\tau_y + 2q_xq_y\tau_x ], \label{eq:quadraticC6}
\end{gather}
where $m_\Delta$ is an effective mass defined by $|\Delta_\bq|= q_{\perp}^2/(2m_\Delta)$. 

In the case of fourfold rotational symmetry, i.e., $n=4$, both $q^2_+$ and $q^2_-$ terms, with angular momenta $l=2$ and $-2$ respectively, are allowed in $\Delta_\bq$. As a result, the Hamiltonian $H(\bq)$ takes a more involved form, which is discussed in the Appendices.

The above cases of chiral pairing with $|l|=1$ and $2$ both give rise to 
gapless Majorana quasiparticles at $\pm \bK$. According to Table \ref{tab:classification} there are two remaining cases which both have $l=0$ mod $n$: $(n,j)=(2,\frac{1}{2})$ and $(6,\frac{3}{2})$. The property $l=0$ mod $n$ implies that spin $\down$ states at $\pm \bK$ are allowed to pair and form a Cooper pair $\Gamma^2_{\bq=0} = c^\dagger_{\bK\down} c^\dagger_{-\bK\down}$ carrying the {\it same} angular momentum $2j=-2j \mod n$ as the spin $\up$ Cooper pair $\Gamma^1_{\bq=0} = c^\dagger_{\bK\up} c^\dagger_{-\bK\up}$. As a result, both Cooper pairs coexist in the superconducting state and generate a full gap at $\pm \bK$.

\begin{figure*}
\includegraphics[width=0.75\textwidth]{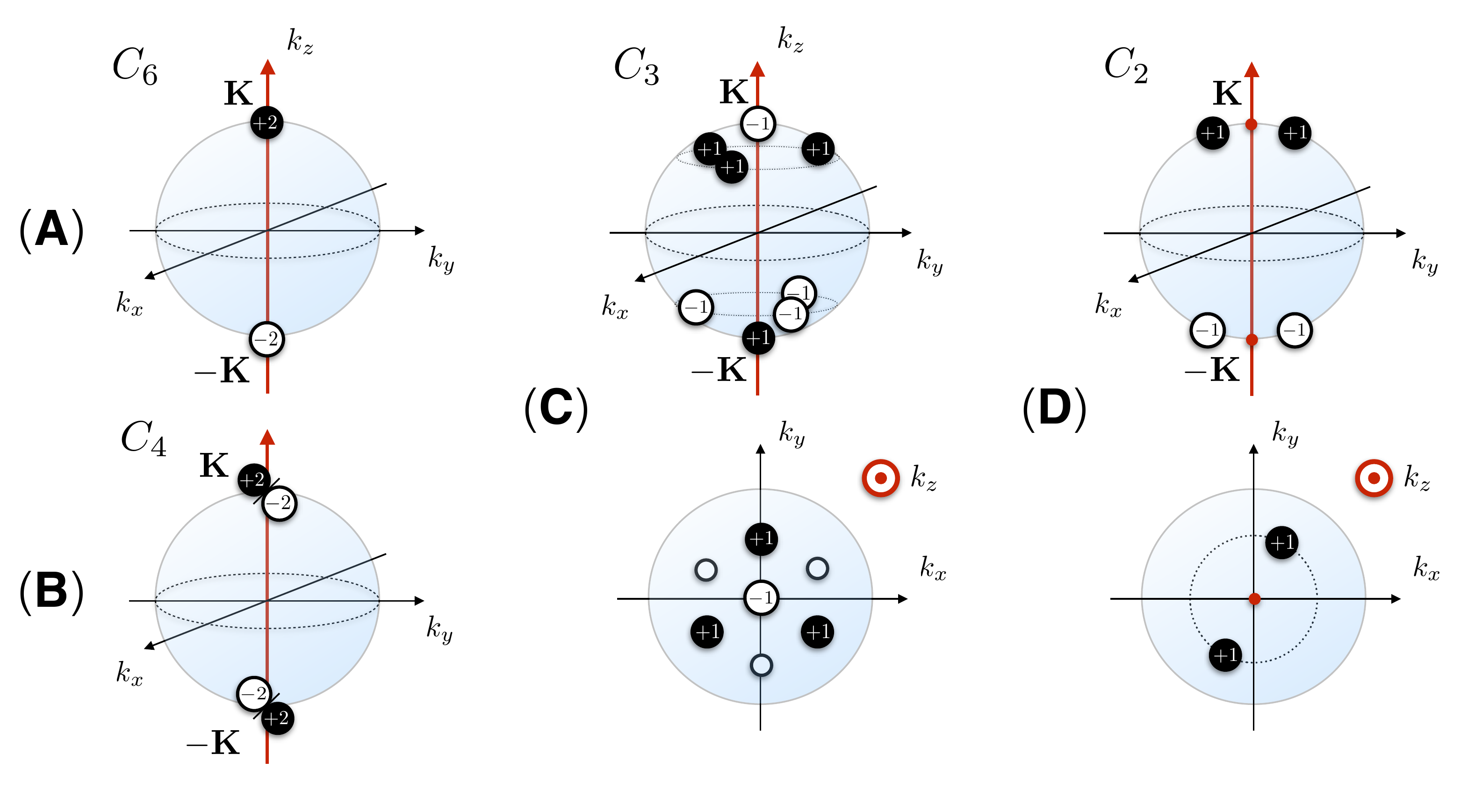}
\caption{\label{fig:nodes} {\bf Schematic structure of Majorana point nodes of spin-orbit coupled chiral superconductors with total angular momentum $J=1$ with an $n$-fold $(n=2,3,4,6)$ rotation axis along $z$.} Two types of Majorana nodes are shown: on-axis and off-axis nodes. Whereas the former are pinned to the rotation axis (i.e., $\pm \bK$), the latter appear at generic Fermi surface momenta. ({\bf  A}) Shows the $C_6$-symmetric case with double Majorana nodes at $\pm \bK$; ({\bf  B}) shows the $C_4$-symmetric case; ({\bf  C} and {\bf D}) show the $C_3$-symmetric and $C_2$-symmetric cases, respectively, including a view from the top (projection on the $xy$ plane). The gap structure of the $C_3$-symmetric superconductor has both on-axis and off-axis nodes, whereas that of the $C_2$-symmetric superconductor only has off-axis nodes. Nodes with positive (negative) monopole charge $C$ [see Eq. \eqref{eq:chern}] are indicated by solid black (white) dots, with the monopole charge (i.e., $C=\pm 1, \pm 2$) explicitly given. In case of $C_4$ symmetry, the sign of the  Majorana node monopole charge at $\pm \bK$ depends on microscopic details (see Appendices).
 }
\end{figure*}


\textbf{\textit{Spin-orbit coupling and non-unitary pairing}}. It is clear from our derivation of the Majorana nodal quasiparticles that these can only be present in chiral superconductors with non-unitary gap structures, i.e., with a spin non-degenerate quasiparticle spectrum such that spin $\up$ and $\down$ states have different gaps \cite{sigristueda}.
Non-unitary superconductors have so far received much less attention than their unitary counterparts. Even though non-unitary pairing states have been discussed in relation to UPt$_3$ \cite{ohmi96,hou98,sauls94,joynt02}, to Sr$_2$RuO$_4$ \cite{sigrist96,machida96,maeno}, and recently to LaNiGa$_2$ \cite{weng16,quintanilla10}, the only established example of non-unitary pairing is superfluid $^3$He in high magnetic fields \cite{ambegaokar73}, known as the A$_1$ phase.
However, from a symmetry point of view, non-unitary pairing is generic and more natural (in a theoretical sense) in chiral superconductors with strong spin-orbit coupling. This is a consequence of the lack of spin-rotational symmetry, replaced by the symmetry of combined spin and momentum rotation under the crystal point group. In such cases, there are typically more than one pairing components with different spin $S$ or orbital angular momentum $L$, but the same {\it total} angular momentum $J=L+S$. As a result, the full bulk gap function $\Delta_\bk$ of the chiral superconductor, defined through $\mathcal{H}_\Delta =   \sum_\bk \;  ( i \Delta_\bk s_y )_{\alpha\beta}c^\dagger_{\bk\alpha}c^\dagger_{-\bk\beta} +\text{H.c.}$, is generally a mixture of these pairing components, all belonging to the same irreducible representation of the point group.
Specifically, $\Delta_\bk$ can we written as
\begin{equation}
\Delta_\bk = \Delta_0 \sum_{t} \lambda_t F^J_t(\bk) , \label{eq:delta_k}
\end{equation}
where $F^J_t(\bk)$ are pairing components (i.e., crystal harmonics) with total angular momentum $J$ but different $L$ and $S$, and $\lambda_t$ are dimensionless coefficients describing the admixture of these different components. For each pairing channel $J$, the set of allowed pairing components $F^J_t(\bk)$ depends both on the point group symmetry of the crystal and the spin angular momentum $j$. In Table \ref{tab:gapfunctions}, we present a full list of gap function components $F^J_t(\bk)$ for trigonal ($C_3$), tetragonal ($C_4$), and hexagonal ($C_6$) superconductors, and for general spin angular momentum $j$. Table \ref{tab:gapfunctions} thus generalizes  standard gap function classifications for $j = \frac{1}{2} $ Bloch electrons \cite{yip93}, and applies to energy bands of, for instance, $j = \frac{3}{2} $ electrons, such as reported in half-Heusler superconductors YPtBi and LuPtBi \cite{brydon16,kim16}.  In addition, the heavy fermion superconductor UPt$_3$ has recently been proposed to have $j=\frac{5}{2}$ bands \cite{nomoto16}.

\begin{table*}[t]
\centering
\begin{ruledtabular}
\begin{tabular}{cccc}
$(J,j)$ & Trigonal ($C_3$)   & Tetragonal ($C_4$)  & Hexagonal ($C_6$)  \\  [4pt]
\hline
$(1,\frac{1}{2})$ & $ k_+s_z, k_zs_+,  k_-s_-,k_zk^2_+s_-,$ & $ k_+s_z, k_zs_+,k_z(k^4_+- k^4_-)s_+,$ & $ k_+s_z, k_zs_+,k_zk^2_+s_-,$    \\[3pt]
 & $ k_zk^2_-s_z, (k^3_+ -k^3_-)s_+$ & $  k_zk^2_+s_-,  k_zk^2_-s_-, k^3_-s_z $  & $ k^5_-s_z,k_zk^4_-s_-,k_zk^6_-s_+$    \\[6pt]
 $(1,\frac{3}{2})$ & $ k_+s_z, k_+s_+, k_+s_-,$ & $ k_+s_z, k_zs_-,k^3_-s_z,$ &  $ k_+s_z, k_zk^2_-s_+,k_zk^2_-s_-,$  \\[3pt]
 & $ k_zk^2_-s_z, k_zk^2_-s_+, k_zk^2_-s_-$ & $ k_zk^2_-s_+, k_zk^2_+s_+,k_zk^4_+s_-$ & $ k^5_-s_z, k_zk^4_+s_+,k_zk^4_+s_-$   \\[6pt]
 $(1,\frac{5}{2})$ & $ k_+s_z,k_-s_+,k_zs_-,$ & $ k_+s_z,k^3_-s_z,k_zk^4_+s_+,$ &  $ k_+s_z,k^5_-s_z,k_zs_-,$  \\[3pt]
 & $ k^3_+s_-,k_zk^2_-s_z,k_zk^2_+s_+$ & $ k_zs_+,k_zk^2_+s_-,k_zk^2_-s_-$ &  $ k_zk^2_+s_+,k_zk^4_-s_+,k_zk^6_+s_-,$  \\[6pt]
$(2,\frac{1}{2})$  & $\cong (-1,\frac{1}{2}) $ & $k_+s_+,k_-s_-,k_zk^2_+s_z,$ & $ k_+s_+,k_zk^2_+s_z,k_zk^4_-s_z$   \\[3pt]
   & & $k^3_+s_-,k^3_-s_+,k_zk^2_-s_z$ & $k^3_+s_-,k^3_-s_-, k^5_-s_+$   \\[6pt]
 $(2,\frac{3}{2})$  & $\cong (-1,\frac{3}{2}) $  &  $k_+s_-,k_zk^2_+s_z,k^3_+s_+,$ &  $k_zk^2_+s_z,k_-s_+,k_-s_-,$  \\[3pt]
 &   &  $k_-s_+,k_zk^2_-s_z,k^3_-s_-$ &  $k_zk^4_-s_z,k^5_+s_+,k^5_+s_- $  \\[6pt]
 $(2,\frac{5}{2})$  & $\cong (-1,\frac{5}{2}) $  &  $k_zk^2_+s_z,k_zk^2_-s_z,k_+s_+,$ &  $k_zk^2_+s_z,k_zk^4_-s_z,k_+s_-,  $  \\[3pt]
 & &  $k^3_-s_+,k_-s_-,k^3_+s_-$ &  $k^5_-s_-,k^3_-s_+,k^3_+s_+  $  \\[6pt]
$(3,\frac{1}{2})$ & $\cong (0,\frac{1}{2}) $ & $\cong (-1,\frac{1}{2}) $ & $k^3_+s_z,k_zk^2_+s_+,k_zk^4_+s_-,$     \\[6pt]
 &  & & $k^3_-s_z,k_zk^2_-s_-,k_zk^4_-s_+$     \\[6pt]
$(3,\frac{3}{2})$ & $\cong (0,\frac{3}{2}) $  & $\cong (-1,\frac{3}{2}) $  &   $k_zs_+,k_zs_-,k_zk^6_+s_+,$    \\[3pt]
 &   &  &  $k^3_+s_z,k^3_- s_z, k_zk^6_-s_-$    \\[6pt]
 $(3,\frac{5}{2})$ & $\cong (0,\frac{5}{2}) $  & $\cong (-1,\frac{5}{2}) $  &   $k^3_+s_z,k^3_-s_z,k_zk^2_-s_+,  $   \\[3pt]
 &   &   &  $k_zk^4_+s_+,k_zk^2_+s_-,k_zk^4_-s_-  $   \\[6pt]
\end{tabular}
\end{ruledtabular}
 \caption{{\bf Complete set of gap functions for chiral spin-orbit coupled superconductors.} Table listing allowed gap function components $F^J_t(\bk)$ of Eq. \eqref{eq:delta_k} for the chiral pairing channels $J=1,2,3$, (pseudo)spin angular momentum $j=\frac{1}{2},\frac{3}{2},\frac{5}{2}$, and crystal rotation symmetries $C_n$ with $n=3,4,6$. For each combination $(J,j)$ a complete set of components is given; any other allowed gap function component $F^J_t(\bk)$ $\Delta_\bk$ is generated by multiplying with fully point group symmetry invariant functions \cite{yip93}. Since angular momenta are only defined mod $n$, some entries in the Table are equivalent, e.g., $(2,\frac{1}{2}) \cong  (-1,\frac{1}{2})$ under $C_3$ symmetry, where $(-1,\frac{1}{2})$ is the time-reversed partner of $(1,\frac{1}{2})$. Recall that $s_\pm = s_x \pm is_y $ and $s_{x,y,z}$ are Pauli matrices acting on the Bloch electron (pseudo)spin.
 }
\label{tab:gapfunctions}
\end{table*}

As an example of $\Delta_\bk$ in Eq. \eqref{eq:delta_k}, consider the following gap function of a $J=1$ superconductor of $j=\frac{1}{2}$ electrons, consisting of two pairing components with $(L,S)=(1,0)$ and $(L,S)=(0,1)$, respectively,
\be
\Delta_\bk =  \lambda_a k_+ s_z + \lambda_b k_z s_+,  \label{eq:deltaC6}
\ee
where we defined $k_\pm = k_x \pm ik_y$ and $s_\pm = s_x \pm i s_y$.
It is straightforward to verify that the pairing is non-unitary as a result of the second $(L,S)=(0,1)$ term. The first term of Eq. \eqref{eq:deltaC6} corresponds to the gap function of the A phase of superfluid $^3$He \cite{leggett75}. The spin-degenerate quasiparticle spectrum of $^3$He A phase is known to host Weyl fermions at low energies \cite{volovikbook,mizushima16}, which can be viewed as a complex quantum field made up of two degenerate Majorana fields. The admixture of the $(L,S)=(0,1)$ component, which is enabled by spin-orbit coupling, gaps out the spin $\up$ states at $\pm \bK$ and gives rise to gapless spin $\down$ excitations governed by Hamiltonian \eqref{eq:quadraticC6} (Appendices).

This example illustrates a general feature of spin-orbit coupled chiral superconductors: lack of spin-rotation symmetry naturally leads to non-unitary pairing, which serves as a spin-selective gapping mechanism and creates spin-non-degenerate nodal excitations which obey the Majorana reality condition.

As we pointed out earlier, the Majorana condition makes the quantum field $\Psi_{\bq}$ a four-component \emph{real} field. One may be tempted to rewrite \eqref{eq:H1} in terms of a two-component complex quantum field $f^\dagger_\bq \equiv(c^\dagger_{\bq 1}, c_{-\bq 2})$: $\mathcal H= f^\dagger_\bq (\xi_\bq \sigma^z + \Delta_\bq \sigma^+ + \Delta_\bq^* \sigma^-) f_\bq$, which is invariant under the $U(1)$ transformation $f^\dagger \rightarrow f^\dagger e^{i \varphi}$.  However, this $U(1)$ symmetry is only present in the presence of translational symmetry and broken by impurity-induced potential scattering between the nodes. To see this, consider the spin-conserving inter-node scattering term
\be
\mathcal H_s  = \sum_\bq M (c^\dagger_{\bq 1} c_{\bq 2} + c^\dagger_{\bq 2} c_{\bq 1} ), \label{eq:mass}
\ee
where $M$ is the scattering amplitude at the momentum $2k_F$. In terms of the complex field $f^\dagger_\bq$, $\mathcal H_s$ involves $f^\dagger_\bq f^\dagger_{-\bq} $ terms, and thus removes the emergent $U(1)$ symmetry in the clean limit. In terms of the four-component Majorana field $\Psi$, $\mathcal H_s$ is given by
\be
 \mathcal H_s  = M \sum_\bq \Psi^\dagger_\bq  \sigma_x\tau_z  \Psi_\bq , \label{eq:mass2}
\ee
Including \eqref{eq:mass} in the Majorana Hamiltonian (\ref{eq:bdg}), the energy $E$ of the Majorana nodes with linear dispersion, Eq. \eqref{eq:linear}, is given by $E^2 = (v_Fq_z)^2 + v^2_\Delta (q_x^2+q_y^2) + M^2. $ 
Thus the effect of inter-node scattering is to generate a mass term without $U(1)$ symmetry, i.e., a Majorana mass term. As a result, the fundamental quantum field describing the gapless quasiparticles is a four-component real field, i.e., a field obeying the Majorana condition.

We note that spin-non-degenerate point nodes also occur when the Fermi surface in the normal state is already spin-split due to magnetism---as theoretically shown in magnetic topological insulator-superconductor heterostructures \cite{meng12} and ferromagnetic $p$-wave superconductors \cite{sau12}, mixing of chiral $d$- and $p$-waves \cite{fischer14}, or due to spin-orbit coupling in noncentrosymmetric superconductors \cite{wang12,sato06,beri10,schnyder11,brydon11,schnyder12,sato11,yada11,tanaka10,sato10}. This is different from our case where the spin-selective point nodes occur via non-unitary pairing in a spin-degenerate normal state. Also, the present case of non-unitary pairing does not assume any special feature in the band structure, and should be distinguished from theoretical surveys of possible pairing states in Weyl and Dirac semimetals \cite{cho12,shivamoggi14,wei14,zhang14,kobayashi15,lu15b,bednik15,li15}.


\textbf{\textit{Off-axis point nodes}}. A key result of our symmetry analysis presented in the first part of this section, is the presence of nodal Majorana excitations at the rotationally invariant Fermi surface momenta $\bK$ on the principal rotation axis for nonzero $l$ mod $n$, see Eq. \eqref{eq:delta_qform}. Rotational symmetry further dictates the form of the energy-momentum dispersion of these \emph{on-axis} Majorana quasiparticles.
Interestingly, we find that spin-orbit coupled odd-parity chiral superconductors can have additional point nodes located at generic Fermi surface momenta away from the north and south poles, i.e., off-axis Majorana nodes. We will first illustrate the presence of these point nodes using examples and then explain their topological origin.

As a first example, let us consider a chiral superconductor with $C_3$ symmetry and angular momentum $J=1$. We now show that its gap structure can exhibit nodes at Fermi surface momenta other than $\pm \bK$. The full pairing potential $\Delta_\bk$ of Eq. \eqref{eq:delta_k} is given to leading $p$-wave order by (see Appendices \ref{app:bcs} and \ref{app:chiral} for details)
\be
\Delta_\bk = \frac{\Delta_0}{k_F} ( \lambda_a k_+ s_z + \lambda_b k_z s_+ + \lambda_c i k_- s_- ), \label{eq:deltaC3}
\ee
where $\lambda_{a,b,c}$ are three real admixture coefficients. In addition to the on-axis nodes at $\bk = \pm \bK$,
the quasiparticle spectrum corresponding to $\Delta_\bk $ of Eq. \eqref{eq:deltaC3} exhibits six nodes
located at off-axis Fermi surface momenta. Writing the Fermi momenta as
$\bk_F = k_F (\cos \varphi_{\bk_F} \sin \theta_{\bk_F},\sin \varphi_{\bk_F} \sin \theta_{\bk_F},\cos \theta_{\bk_F})$,
the location of these nodes can be expressed as the relations
$\cos \theta_{\bk_F} = \pm\lambda_a^2/\sqrt{\lambda_a^4+16\lambda_b^2\lambda_c^2}$
and $\sin 3\varphi_{\bk_F} = \mp 1$ \cite{note1}. There are three nodes on the northern Fermi surface
hemisphere, which are related by threefold rotation $C_3$, and each of these nodes has a partner on
the southern hemisphere related by inversion $P$, as shown in Fig. \ref{fig:nodes}(c).

As a second example, consider a $J=1$ superconductor with twofold rotational symmetry $C_2$. The pairing potential $\Delta_\bk $ is composed of all terms which are odd under $C_2$ and to $p$-wave order in spherical harmonics is given by
\be
\Delta_\bk = \frac{\Delta_0}{k_F} [ ( \lambda_a k_+ + \lambda_b k_-) s_z + k_z (\lambda_c s_+ + \lambda_d s_-) ].\label{eq:deltaC2}
\ee
At the Fermi surface momentum $\bK$ one has $\Delta_\bK = \Delta_0(\lambda_c s_+ + \lambda_d s_-)$, which implies a pairing gap for both $c_{\pm\bK +\bq \up}$ and $c_{\pm\bK +\bq \down}$, in agreement with the result of Table \ref{tab:classification}. Even though no nodes exist on the twofold $z$-axis, it is straightforward to verify that for general nonzero admixture coefficients two pairs of nodes are located on the Fermi surface, each pair related by twofold rotation with the partners of a pair related by the inversion $P$. Taking $\lambda_b=0$ ($\lambda_a=0$), for instance, the nodes are located on the intersection of the Fermi surface with the $yz$ ($xz$) plane, as shown in Fig. \ref{fig:nodes}(d).

In both these examples, all point nodes located at generic rotation non-invariant Fermi surface momenta are spin-non-degenerate, and the gapless quasiparticles are therefore Majorana fermions. We call these nodes \emph{off-axis} Majorana nodes. The presence of the off-axis Majorana nodes in these examples motivates the question whether these are accidental, or whether the off-axis nodes are related to the Majorana nodes on the rotation axis at a deep level.

We now address this question by focusing on the topological nature of the Majorana point nodes. We will show that the classification of different types of low-energy Majorana quasiparticles in terms of location on the Fermi surface (i.e., on-axis or off-axis) and energy-momentum dispersion (i.e., linear or quadratic) is linked to a topological property of point nodes in momentum space.

In band theory crossing points of (quasiparticle) energy bands are endowed with an integer topological quantum number given by the Chern number $C$ defined as
\beq
C = \frac{1}{4\pi}\int_\Omega \mathcal{F}\cdot d\bS. \label{eq:chern}
\eeq
Here $\mathcal{F} = \boldsymbol{\nabla}_\bk \times  \mathcal{A}(\bk)$ is the Berry curvature, i.e., the field strength of the momentum space gauge Berry connection $\mathcal{A}(\bk)$, which is integrated over a surface $\Omega$ enclosing the point node. Hence, $C$ quantifies the Berry curvature monopole strength of the point node. Since monopoles cannot be removed unless they annihilate with a monopole of equal but opposite strength, point nodes which carry a nonzero monopole charge are topologically protected.

It is straightforward to show that the Majorana nodes with linear dispersion described by Eq. \eqref{eq:linear} have monopole charge $C=\mp1$ at $\pm \bK$ similar to Weyl fermions in topological semimetals \cite{wan11,turner13}, and that the Majorana nodes of Eq. \eqref{eq:quadraticC6}, which disperse quadratically tangential to the Fermi surface, have monopole charge $C=\pm 2$, similar to double-Weyl fermions  \cite{fang12}. Therefore, the former may be called \emph{single} Majorana nodes and the latter \emph{double} Majorana nodes. These single and double on-axis Majorana nodes, corresponding to $C_3$ and $C_{4,6}$ symmetry, respectively, are schematically shown in Fig. \ref{fig:nodes}, with the monopole charge $C$ explicitly indicated.

From the perspective of topology, the presence of the off-axis Majorana nodes in superconductors with $C_3$ or $C_2$ can be understood from monopole charge conservation. For instance, the $C_3$-symmetric superconductor with the gap function \eqref{eq:deltaC3} can be viewed as a descendent of the $C_6$-symmetric superconductor with the gap function \eqref{eq:deltaC6}, obtained from lowering the symmetry. As a result of lower symmetry, additional gap functions can mix in with the $J=1$ channel, and according to our symmetry analysis this transforms the $C=2$ double Majorana node at $\bK$ of a $C_6$ superconductor into $C=-1$ Majorana node at $\bK$ of a $C_3$ superconductor. Since the total monopole charge must be conserved, and the gap structure must preserve $C_3$ symmetry, there must exist three additional point nodes with monopole charge of $C=+1$. This effective ``splitting'' of a $C=2$ Majorana node into four single nodes is in agreement with the explicit analysis of Eq. \eqref{eq:deltaC3} and is schematically shown in Fig. \ref{fig:nodes}. An analogous argument explains the existence of two $C=+1$ off-axis Majorana nodes in case of the $C_2$-symmetric superconductor, which has a full pairing gap at $\bK$. The nodal structure of the  $C_2$ superconductor is depicted in Fig. \ref{fig:nodes}. It is therefore natural to think of the off-axis Majorana nodes as originating from  on-axis double Majorana nodes in a $J=1$ superconductor, obtained by lowering $C_6$ or full rotational symmetry to trigonal ($C_3$) and orthorhombic ($C_2$) symmetry.


These topological arguments demonstrate that the chiral superconductors discussed here are topological nodal superconductors. Topological nodal superconductors are superconductors with topologically protected gapless quasiparticle excitations, in a way which parallels the protection of Weyl fermions in Weyl semimetals \cite{wan11,turner13}  (see also Section ``Surface Andreev bound states: Majorana arcs''). In particular, the gapless Majorana quasiparticles are topologically protected by monopole charge conservation.

\subsection*{Detecting 3D Majorana fermions \label{sec:detect}}

In this section we explore ways to detect the 3D Majorana nodal quasiparticles in chiral spin-orbit coupled superconductors. Since the Majorana fermions arise as a result of a spin-selective gapping mechanism associated with non-unitary pairing, we first address the experimentally observable consequences of non-unitary pairing from a general perspective, and then turn to a specific probe sensitive to the spin-polarized low-energy Majorana quasiparticles: the NMR spin relaxation rate.


\textbf{\textit{Signatures of non-unitary chiral superconductors}}. The non-unitary gap structure of chiral spin-orbit coupled superconductors gives rise to a number of distinctive experimental signatures.  
The most prominent characteristic of non-unitary pairing, the non-degenerate quasiparticle excitation spectrum, leads to a spin-dependent density of states: the density of states $N_{\up,\down}(E)$ for (pseudo)spin-$\up,\down$ excitations are unequal, i.e., $N_\up(E) \neq  N_\down(E)$, and in particular, low-energy branch of the spectrum consists of fully spin-polarized states near the point nodes. As a consequence of the non-degenerate spectrum, the total density of states $\sum_\alpha N_\alpha(E)$ can exhibit two distinct peaks, rather than a single peak at $\Delta_0$ which is characteristic of conventional $s$-wave superconductors. This can lead to a two-gap like feature in the specific heat, as is demonstrated more explicitly on the basis of simple examples in the Appendices. Therefore, experimental signatures that are commonly attributed to multiband superconductivity may actually originate from non-unitary pairing in a single band.

A well-known and discriminating property of nodal superconductors is the characteristic temperature dependence of a diverse set of dynamic and thermodynamic quantities such as the electronic part of the specific heat, the London penetration depth, and the NMR spin relaxation rate $T^{-1}_1$ \cite{matsuda06}. The low-energy branch of the quasiparticle spectrum of chiral non-unitary superconductors consists of $C=\pm 1$ or $C=\pm 2$ point nodes, and this implies that the density of states $N(E)$ in the regime $E \ll \Delta_0$ takes the form $N(E)/N_0 \sim (E/\Delta_0)^{n}$, where $N_0$ is the normal state density of states. The exponent $n$ depends on the nature of the point node: it equals $n=2$ for $C=\pm 1$ and $n=1$ for $C=\pm 2$. 
The form of the density of states at the nodes is responsible for the typical power law temperature dependence of the specific heat, the penetration depth, and the spin relaxation, which probe the density of quasiparticle states, at temperatures $T \ll T_c \sim \Delta_0$.

In the next section we consider the NMR spin relaxation rate in more detail. In case of chiral superconductors with low-energy Majorana quasiparticles, not only should the spin relaxation rate $T^{-1}_1$ exhibit power law temperature dependence at low temperatures, the temperature dependence of $T^{-1}_1$ is expected to crucially depend on the direction of the nuclear spin polarization. This follows from the fact that the only quasiparticle excitations available at low-energy are spin-$\down$ states, which is intimately related to the non-unitary nature of the superconducting state. Therefore, we derive below theory of NMR spin relaxation in non-unitary superconductors, which serves as a powerful tool to identify Majorana nodal quasiparticles.


\textbf{\textit{NMR: spin relaxation rate}}. The measurement of the NMR spin-lattice relaxation rate $1/T_1$ at low temperature is a well-established experimental technique to probe to the gap structure of superconductors. The temperature dependence of $1/T_1$ at temperatures $T \ll T_c \sim \Delta_0$ can be used as a measure of the density of low-energy quasiparticle states, and allows to distinguish fully gapped, point nodal, and line nodal gap structures \cite{sigristueda}.

The coupling of quasiparticle states to the nuclear spin originates from the hyperfine interaction between the nuclear spin and itinerant electrons. The hyperfine coupling Hamiltonian $\mathcal H_{\text{hf}}$ is given by 
\begin{equation}
\mathcal H_{\text{hf}} = \gamma_N A_{\text{hf}} \sum_{\bk \bk'} g_{ij}(\bk,\bk')\hat{S}^i c_{\bk \alpha}^\dagger s^j_{\alpha\beta} c_{\bk' \beta}, \label{eq:Hhfa}
\end{equation}
where $\hat{S}^i $ are the components of the nuclear spin operator, $s^i$ are Pauli matrices representing the electron pseudospin (summation of repeated spin indices $\alpha,\beta$ implied). Furthermore, $\gamma_N$ is the gyromagnetic ratio of the nuclear spin, $A_{\text{hf}}$ is the hyperfine coupling constant, and $g_{ij}(\bk,\bk')$ is a momentum-dependent tensor describing the coupling of the nuclear spin to the pseudospin of electrons in spin-orbit-coupled materials.
The form of this tensor can be complicated and material-specific. However, since only quasiparticles around the nodes contribute to the spin relaxation at low temperature, it suffices  to consider $g_{ij}(\bk,\bk')$ at the nodes, a key simplification that enables us to find universal features of the spin relaxation rate below.

Our aim is to derive $1/T_1$ for chiral non-unitary superconductors hosting 3D Majorana fermions. Since the Majorana nodes are non-degenerate with definite pseudospin, the low-energy quasiparticles couple anisotropically to the nuclear spin. To demonstrate this, consider the case of gapless Majorana particles pinned to $\pm \bK$. Projecting the Hamiltonian \eqref{eq:Hhfa} into the space of Bloch states near $\pm \bK$ ($\mathcal{P}$ projects onto the low-energy Hilbert space) one finds
\begin{multline}
\mathcal{P}\mathcal H_{\text{hf}} \mathcal{P} = \gamma_N A_{\text{hf}} \sum_{\bq\bq'} \hat S^z  \left[ g^1_{zz} c^\dagger_{\bq 1} c_{\bq' 1} + g^2_{zz}  c^\dagger_{\bq 2} c_{\bq' 2}  \right. \\
\left.   +  g^3_{zz} c^\dagger_{\bq 1} c_{\bq' 2} + g^{3*}_{zz} c^\dagger_{\bq 2} c_{\bq' 1}  \right], \label{eq:Hhflowenergy}
\end{multline}
where $g^1 = g(\bK, \bK)$, $g^2=g(-\bK, -\bK)$, and $g^3=g(\bK, -\bK)$ are the $g$-tensors evaluated at the nodes. Importantly, only the $z$-component of the nuclear spin enters the low-energy Hamiltonian due to the rotational symmetry of the crystal around the $z$ axis. As a result, we expect that the nuclear spin relaxation time $1/T_1$ is highly direction dependent, and to leading order approximation, diverges when the nuclear spin is initially polarized along the $z$ direction. 

As we will show below, the strongly anisotropic spin relaxation rate is a generic consequence of spin-selective nodes, whereas the divergence for nuclear spin polarization along the nodal direction is an artifact of the leading-order Hamiltonian (\ref{eq:Hhflowenergy}). We now go beyond this leading-order approximation and  expand the full form factor $g_{ij}(\bk,\bk')$ into crystal spherical harmonics. Keeping the lowest-order $s$-wave component, we have $g_{ij}(\bk,\bk') \sim \delta_{ij}$, reducing the hyperfine coupling to
\be
\mathcal H_{\text{hf}} = \gamma_N A_{\text{hf}} \sum_{\bk \bk'} \hat{S}^i c_{\bk \alpha}^\dagger s^i_{\alpha\beta} c_{\bk' \beta}. \label{eq:Hhfb}
\ee

With Eq. \eqref{eq:Hhfb} we proceed to calculate the NMR relaxation rate $1/T_1$ for a non-unitary superconductor explicitly. For simplicity we consider a nuclear spin of $S=1/2$. The spin relaxation rate $1/T_1$ is expressed through the transverse spin susceptibility $\chi^{-+}(\bp,\omega)$ (i.e., transverse to nuclear spin direction) and reads as \cite{hebel59,moriya}
\be
\frac{1}{T_1} = \gamma_N^2 {A_{\text{hf}}}^2 T \lim_{\omega\to 0}\sum_{\bp} \frac{\text{Im} \chi^{-+}(\bp,\omega)}{\omega}. \label{eq:T1}
\ee
When evaluating $1/T_1$ it is important to distinguish unitary and non-unitary pairing states. In case of the former, it can be shown that the spin relaxation rate does not depend on the direction of polarization of the nuclear spin. (This result is reviewed in the Appendices.) The case of non-unitary pairing requires separate and more careful treatment. For concreteness, here we first consider the $C_6$-symmetric $J=1$ superconductor, which only has double nodes at $\pm \bK$. Other nodal non-unitary superconductors are discussed towards the end of this section.

The gap structure of the $J=1$ superconductor with $C_6$ symmetry is given by Eq. \eqref{eq:deltaC6}. The full BdG Hamiltonian is diagonalized in terms of Bogoliubov quasiparticle operators which we define as $a_{\bk \alpha}$. Writing the hyperfine interaction Hamiltonian \eqref{eq:Hhfb} in terms of the Bogoliubov quasiparticle operators, the spin susceptibility can be readily evaluated. Since at low temperatures (i.e., $\sim T\ll \Delta_0 \min\{\lambda_a, \lambda_b\}$) only low-energy excitations contribute to the relaxation rate, we can restrict to Bogoliubov quasiparticle states with small momenta $\bq$ relative to $\pm \bK$ and only keep the low-energy gapless states $a_{\bq}$ corresponding to energies $E_{\bq} = [\xi_{\bq}^2 + (q_{\perp}^2/2m_\Delta)^2]^{1/2}$ [see Eq. \eqref{eq:quadraticC6}], where $q_{\perp} =(q_x^2+q_y^2)^{1/2}$. Then, to the leading order in small momentum $\bq$, the Hamiltonian \eqref{eq:Hhfb} reads as
\begin{multline}
\mathcal H_{\text{hf}} \cong \gamma_N A_{\text{hf}} \sum_{\bq \bq' pp'} a_{\bq p}^\dagger a_{\bq' p'} \left[ \hat S_z F_{pp'}(\bq,\bq')+ \right. \\ \left. p' \hat S_- G(\bq,\bq') + p \hat S_+  G^*(\bq',\bq) \right],   \label{eq:Hhfnodes}
\end{multline}
with the form factors $F_{pp'}(\bq,\bq') $ and $G(\bq,\bq')$ defined as
\beq
F_{pp'}(\bq,\bq') &=& \frac1{2q_{\perp}q_{\perp}'}\left(q_+q_-'P_--q_-q_+'pp'P_+ \right), \nonumber \\
G(\bq,\bq') &=& \frac{1}{\sqrt{8m_\Delta \tilde \Delta_0}} \left(\frac{q_{\perp} q_{+}'}{q_{\perp}'}P_+ - \frac{ q_+q_{\perp}'}{q_{\perp}}P_- \right),  \label{eq:factors}
\eeq
and the momentum-dependent factors $P_{\pm} = \sqrt{(1\pm \xi_{\bq}/E_{\bq})(1\pm \xi_{\bq'}/E_{\bq'})}$. Here $p,p' = +1$ ($-1$) for the north (south) node.

In Eq. \eqref{eq:Hhfnodes} the anomalous terms $a_{\bq} a_{\bq'}$ and $a^\dagger_{\bq} a^\dagger_{\bq'}$ have been omitted, since they do not contribute to $1/T_1$ due to energy conservation. The effective mass $m_\Delta$ of the the Bloch states $c_{\pm \bK\down}$ and energy $\tilde \Delta_0$ associated with the Bloch states $c_{\pm \bK\up}$ are defined in terms of gap function parameters in the Appendices. Equation \eqref{eq:Hhfnodes}, which is the projection of Hamiltonian \eqref{eq:Hhfa} into the space of low-energy Bogoliubov quasiparticles, should be compared to the projection into the space of low-energy Bloch states given in Eq. \eqref{eq:Hhflowenergy}. The former contains a coupling to $\hat{S}_\pm = \hat{S}_x \pm i  \hat{S}_y$ originating from the nonzero support of the Bogoliubov quasiparticle states on the Bloch electron states $c_{\pm \bK+ \bq \up}$. The support is vanishingly small near the nodes as a result of $ q_{\pm}/(m_\Delta\tilde{\Delta}_0)^{1/2}\propto q_{\pm}/\kf \ll 1$, indicating a suppression of the spin relaxation rate for a nuclear spin initially polarized along the $z$-direction. In the limit that such coupling is strictly absent, as in Eq. \eqref{eq:Hhflowenergy}, spin initially polarized along the $z$-direction does not relax.

Using the projection of the hyperfine coupling into the space of low-energy Bogoliubov quasiparticles given by \eqref{eq:Hhfnodes} it is straightforward to calculate NMR relaxation rate given by Eq. \eqref{eq:T1} in the low temperatures limit $T\ll \Delta_0 \min\{\lambda_a, \lambda_b \}$ (Appendices). We find $1/T_1$ to be given by
\begin{equation}
\frac1{T_1} = D_{\perp} T^3 S_{\perp}^2 + D_z  \frac{T^4}{\tilde \Delta_0} S_z^2,  \label{eq:NMRrate}
\end{equation}
where $S_{\perp}$ and $S_{z}$ are the projections of the nuclear spin polarization on the $xy$ plane and $z$ axis, respectively, and the coefficients $D_{\perp,z}$ are given by
\begin{gather}
D_{\perp} = \gamma_N^2 A^2_{\text{hf}}\frac{\pi }{ 96} \left(\frac{ m_{\Delta}}{\vf}\right)^2, \quad D_z =  \gamma_N^2 A^2_{\text{hf}} \frac{9\zeta(3)}{4\pi^2 } \left(\frac{ m_{\Delta}}{\vf}\right)^2 \label{eq:NMRcoeff}
\end{gather}
Equation \eqref{eq:NMRrate} proves that there is a strong anisotropy of spin relaxation depending on the polarization of the nuclear spin. For a nuclear spin polarized perpendicular to the $z$ axis, the spin relaxation behaves as $1/T_1 \sim T^3$, as expected for $C=\pm 2$ point nodes with quadratic dispersion and linear dependence of density of states on energy. Instead, for a nuclear spin polarized along the $z$ axis, the spin relaxation rate is suppressed by a factor of $T/ \tilde \Delta_0$, i.e., the zeroth order term in an expansion in $T/ \tilde \Delta_0$ is absent.

The strong relaxation rate anisotropy can be intuitively understood from a simple physical picture. The hyperfine interaction leading to nuclear spin relaxation is given by Eq. \eqref{eq:Hhfb}, and consequently, nuclear spin relaxation occurs simultaneously with an electron spin flip. At low-energies, however, only $c_{\pm \bK + \bq \down}$ Bloch states are available, implying that a nuclear spin initially polarized along $z$ is vanishingly improbable to relax. This physical picture is captured by Eq. \eqref{eq:Hhflowenergy}. The qualitative difference of this result compared to the case of unitary pairing \cite{sigristueda} can be understood in a similar way. In the case of the latter, the quasiparticle energy spectrum is doubly degenerate, implying that both spin species are present and, consequently, leading to the same temperature dependence of $1/T_1$ for all nuclear spin polarizations.

A qualitatively similar anisotropic spin relaxation rate has been predicted for two-dimensional Majorana fermions, which live on the surface of a 3D topological superfluid, i.e., the $^3$He-B phase~\cite{chung09,nagato09}. In that case, where the 2D surface Majorana modes are described by a two-component real quantum field, the anisotropy in the spin relaxation arises since one can only construct Ising spin operator, which points perpendicular to the surface and takes the form of a Majorana mass term. This is different from our case, i.e., the 3D Majorana fermions, where the spin operator constructed from the low-energy quasiparticles does not correspond to a mass term.

Based on the result for the $J=1$ superconductor with $C_6$ symmetry and quadratic point nodes at $\pm \bK$, Eq. \eqref{eq:NMRrate}, we now comment on other chiral non-unitary superconductors. First, consider the $J=1$ superconductor with $C_4$ symmetry shown in Fig. \ref{fig:nodes} and discussed in more detail in Appendix \ref{app:symmetry}. Since the $C_4$-symmetric superconductor only has on-axis Majorana nodes with $C= \pm 2$, Eq. \eqref{eq:NMRrate} remains valid and a significant suppression of $1/T_1$ for the nuclear spin polarized along $z$-axis is expected.

Next, consider the $J=1$ superconductor with $C_3$ symmetry with gap function \eqref{eq:deltaC3}. In this case, the low-energy gap structure consists of eight single Majorana nodes (i.e., linear dispersion), two of which located at $\pm \bK$, and six at off-axis Fermi surface momenta (see Fig. \ref{fig:nodes}). Such nodal structure complicates the explicit derivation of an analytical expression for $1/T_1$, but the final result, however, can be inferred from the $C_6$ symmetric case. For a weak trigonal anisotropy and intermediate temperatures given by the condition $\Delta_0\lambda_c<T<\Delta_0\min\{\lambda_a,\lambda_b  \}$, the trigonal $\lambda_c$-term can be neglected and one can expect the same behavior as in the hexagonal case with the relaxation rate given by Eq. \eqref{eq:NMRrate}.

At the lowest temperatures given by $T\ll\Delta_0\min\{ \lambda_a,\lambda_b ,\lambda_c \}$, however, the linear dispersion of the nodes comes into play: the density of states becomes a quadratic function of energy, leading to $1/T_1\sim T^5$. Whereas at the on-axis nodes still only quasiparticles with definite spin $\down$ are available at low-energies, the low-energy quasiparticles at the off-axis nodes are mixtures of spin $\up$ and $\down$. As a result, the temperature dependence of $1/T_1$ will exhibit the same power law behavior, i.e., $\sim T^5$ for all nuclear spin polarizations. The numerical prefactors will reflect a directional anisotropy, however, which may be large.

Superconductors with $C_2$ symmetry, due to the four off-axis linear nodes, are expected to show the behavior similar to $C_3$ crystals, exhibiting strong anisotropy in the functional temperature dependence at intermediate temperatures and point node power law behavior at the lowest temperatures for all nuclear spin polarizations. Again, numerical prefactors will generically be different.

\subsection*{Surface Andreev bound states: Majorana arcs \label{sec:arcs}}

\begin{figure}
\includegraphics[width=\columnwidth]{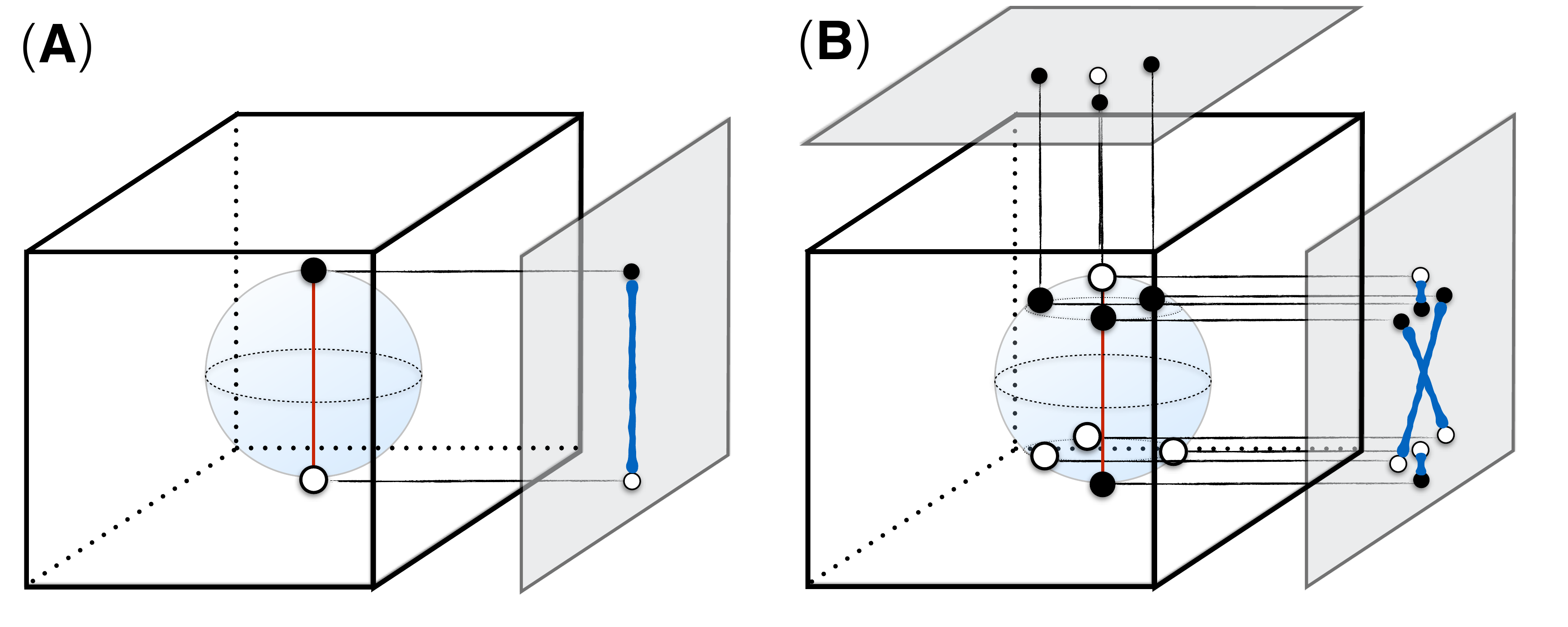}
\caption{
{\bf Schematic representation of arc surface Andreev bound states of nodal superconductors.} For a given surface termination, the projections of the bulk Majorana nodes onto the surface momentum space (transparent gray planes) are connected by the surface Majorana arcs (thick blue lines). The surface Majorana arcs must start and terminate at nodes with opposite monopole charge. ({\bf A}) Arc structure on a side surface of the A phase of $^3$He. ({\bf B}) Schematic arc structure of $C_3$-symmetric $J=1$ chiral superconductor for a side surface in the $y$ direction (see also Fig. \ref{fig:arcs}). The projection of bulk Majorana nodes (coming from northern Fermi surface hemisphere) on the top surface is also shown, cf. Fig. \ref{fig:arcs}({\bf B}).}
\label{fig:schematicarcs}
\end{figure}

Chiral superconductors with point nodal quasiparticle excitations are topological nodal superconductors due to the nonzero Berry monopole charge of the point nodes \cite{matsuura13}, as discussed following Eq. \eqref{eq:chern}. Via the bulk-boundary correspondence, the topological nature of the bulk superconducting state is reflected on a surface boundary separating the superconductor from the vacuum or, equivalently, a gapped $s$-wave superconductor. The surface Andreev bound states of topological nodal superconductors take the form of arcs in surface momentum space, connecting the projections of the topological bulk nodes onto surface momentum space, schematically shown in Fig. \ref{fig:schematicarcs}. A canonical example of surface states in nodal topological systems are the surface Fermi arcs of the superfluid $^3$He A phase, which originate from and terminate at projections of the bulk Weyl nodes \cite{tsutsumi11,silaev12}. A similar surface state structure has been explored in the context of UPt$_3$ \cite{goswami15}.

Weyl nodal fermions have recently attracted a great deal of attention in semimetallic materials, referred to as Weyl semimetals \cite{wan11,burkov11,burkov11b,hosur12,turner13}, which are semimetals with non-degenerate point nodal touchings of bulk energy bands, associated with nonzero Berry monopole charge, and Fermi arcs on the surface. The Weyl semimetal state has recently been predicted and observed in the TaAs materials class \cite{weng15,huang15,xu15,xu15b,lv15,lv15b,yang15,bernevig15}, and photonic crystals \cite{lu13,lu14,lu15}.
The superconducting analog of Weyl semimetals was first considered in Ref. \onlinecite{meng12} based on a topological insulator-$s$-wave-superconductor heterostructure model. The surface arcs connect the projections of the non-degenerate bulk nodes in the Bogoliubov quasiparticle spectrum, and due to the redundancy built into the BdG mean-field description these surface arcs are Majorana arcs.

In general, non-degenerate nodal touchings of energy bands can only occur when at least one of two symmetries, time-reversal symmetry $\Theta$ or inversion symmetry $P$, is broken \cite{young12}. The experimentally found Weyl semimetals in TaAs and related materials all break inversion symmetry, and even though much effort has been devoted to looking for time-reversal breaking Weyl semimetals \cite{wan11,burkov11,burkov11b,xu11,yang11,fang12,krempa12,kim13,bulmash14,dubcek15,borisenko15,chang16}, their conclusive observation in materials remains an open challenge.

The chiral superconductors of this work break time-reversal symmetry and have odd-parity pairing, implying that they preserve a $Z_2$ symmetry given by $\tau_z P$ (see Materials and Methods). Hence, the only symmetry manifest at any surface is particle-hole symmetry. Since the bulk point nodes are Majorana nodes, the surface states of chiral non-unitary superconductors are Majorana arcs.

In this section we calculate and study the structure of the Majorana arcs of chiral non-unitary superconductors, with a focus on $J=1$ superconductors with $C_6$ and $C_3$ symmetry with gap functions \eqref{eq:deltaC6} and \eqref{eq:deltaC3}, respectively. The profile of surface Majorana arcs depends on the projections of the bulk Majorana nodes onto the surface, and therefore depends on boundary geometry. Here we will consider two different semi-infinite geometries: (i) a boundary in the $xz$ plane, separating the vacuum ($y<0$) and the superconductor ($y>0$), and (ii) a boundary in the $xy$ plane (i.e., vacuum $z<0$ and superconductor $z>0$).

\begin{figure}
\includegraphics[width=0.9\columnwidth]{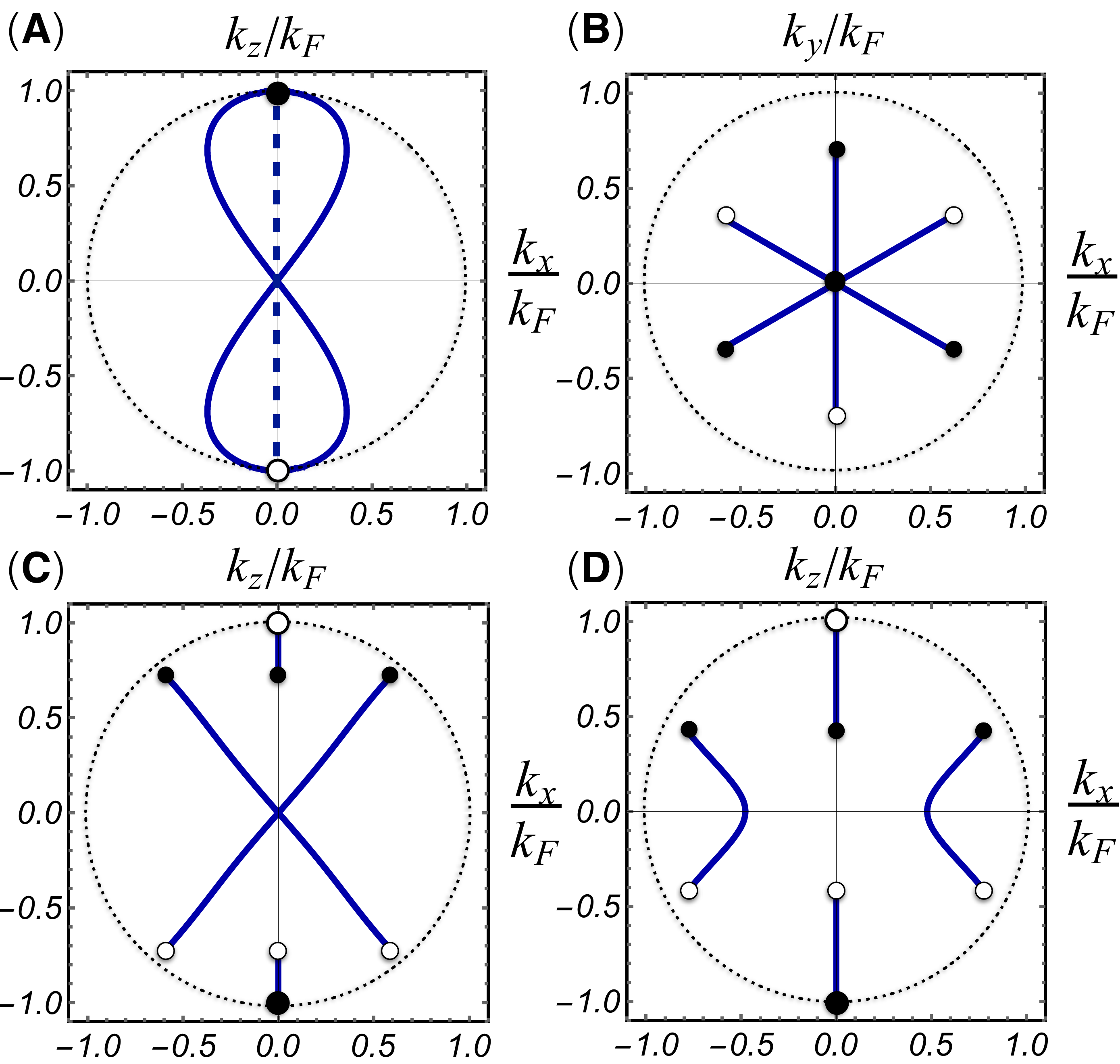}
\caption{ {\bf Majorana arc surface states.} 
Plots of the zero-energy ($E=0$) surface Majorana arc states in surface momentum space for chiral $J=1$ superconductors with $C_6$ symmetry ({\bf A}) and $C_3$ symmetry ({\bf B}--{\bf D}), and gap functions \eqref{eq:deltaC6} and \eqref{eq:deltaC3}, respectively. Panels ({\bf A}), ({\bf C}), ({\bf D}) show the Majorana arc states of a surface boundary in the $xz$ plane, i.e., semi-infinite superconductor at $y>0$, whereas ({\bf B}) shows the Majorana arc states of surface boundary in the $xy$ plane (superconductor $z>0$). As all panels show, the surface Majorana arcs connect the projections of the bulk Majorana nodes. The dashed circle shows the radius of the Fermi surface projection. In panel ({\bf A}) the straight dashed blue line denotes the Fermi arcs of superfluid $^3$He-A for comparison. The parameters used are given by $\lambda_a\Delta_0/\mu = 0.013$, $\lambda_b\Delta_0/\mu = 0.01$, and $\lambda_c\Delta_0/\mu = 0.004,0.009$ in ({\bf B}, {\bf C}) and ({\bf D}), respectively.
}
\label{fig:arcs}
\end{figure}

Starting with a boundary in the $xz$ plane, the first quantized Hamiltonian is given by $H_{\text{BdG}}(\bk,\br) = H_{\text{BdG}}(\bk)\theta(y)$ with $H_{\text{BdG}}(\bk)$ the (first quantized) BCS-BdG Hamiltonian of the superconductor with pairing potential \eqref{eq:deltaC6}. We solve the equation $H_{\text{BdG}}(-i\boldsymbol{\nabla},\br) \Psi(\br) = E \Psi(\br)$ for zero-energy solutions $E=0$, localized at the boundary, i.e., wave functions decaying exponentially at $y\to \infty$ and satisfying the Dirichlet boundary condition $\left.\Psi\right\vert_{y=0}=0.$ In this geometry, $k_x$ and $k_z$ remain good quantum numbers and we substitute $k_y\to -i\partial_y$ in $H_{\text{BdG}}(\bk)$.

The gap function of a $C_6$-symmetric superconductor, given by Eq. \eqref{eq:deltaC6}, gives rise to non-degenerate $C= \pm 2$ bulk nodes at $\pm \bK$. To solve for the zero energy states of the Majorana arcs, we look for a general solution of the form $\Psi(y) \propto \exp(\alpha y)$ with the condition $\text{Re}\, \alpha <0.$ The equation $H_{\text{BdG}}(-i\nabla_y)\Psi=0$ then translates into a polynomial in $\alpha$ of degree eight, whose roots determine the wave function solutions. The roots can be found explicitly and are given by
\begin{multline}
 \alpha_{3(4)} = \alpha_{1(2)}^*, \qquad \alpha_{1,2}^2 = -k_\perp^2 + 2\tilde\lambda_a^2+2i\tilde\lambda_bk_z  \\
\pm 2\sqrt{(\tilde\lambda_a^2+i\tilde\lambda_bk_z)^2 - \tilde\lambda_a^2(\kf^2-k_z^2)}, \label{eq:arcsC6}
\end{multline}
where we defined $k_\perp \equiv\sqrt{\kf^2-k_x^2-k_z^2},$ $\tilde \lambda_{a(b)} \equiv \Delta_0  \lambda_{a(b)}/\vf,$ and the eigenvalues with $\text{Re} \, \alpha_{1,2}<0$ are implied, as only they can be used to satisfy the localization condition $\left.\Psi\right\vert_{y\to +\infty}\to 0$. These four eigenvalues are then used to construct the solution that satisfies the Dirichlet boundary condition at $y=0$, $\left.\Psi\right\vert_{y=0}=0.$ This condition results in the implicit equation for the zero energy Majorana arc profile in the $k_x-k_z$ plane. The implicit equation reads as
\be
|k_x(\alpha_1+\alpha_2)| = |k_\perp^2-\alpha_1\alpha_2|.
\ee
The resulting profile of (zero energy) Majorana arc states has a ``bow-tie'' shape in surface momentum space and is shown in Fig. \ref{fig:arcs}(a). The zero energy solutions are non-degenerate, apart from the surface momentum $(k_x,k_z)=(0,0)$, but are related by particle-hole symmetry. Such profile should be compared to the fully degenerate surface Majorana arcs of superfluid $^3\text{He}$-$A$, corresponding to $\lambda_b=0$ in Eq. \eqref{eq:deltaC6} and shown with a dashed line in Fig. \ref{fig:arcs}(a) (see also Fig. \ref{fig:schematicarcs}). As a result of the degeneracy of the two sheets of Majorana arcs they effectively form a single complex Fermi arc.

In the vicinity of $(k_x,k_z)=(0,0)$, the structure of the Majorana arcs of the $C_6$-symmetric chiral superconductor can be obtained within the semiclassical or Andreev approximation (we follow Ref. \onlinecite{park15}). Defining $k_\perp = \sqrt{\kf^2-k_x^2-k_z^2}$ as the semiclassical momentum perpendicular to the boundary and assuming $k_\perp \gg k_F (\Delta_0/\varepsilon_F)$, one obtains the Andreev Hamiltonian from the BdG Hamiltonian as $H_{\text{BdG}}(\bk,\br) \rightarrow H_{\perp}(k_\perp,-i\nabla_y) + H_{\parallel}(\bk_\parallel)$, with $\bk_\parallel \equiv (k_x,k_z)$ and $H_{\perp,\parallel}$ (Appendices). In obtaining the surface Majorana arc Hamiltonian, one first solves $H_{\perp}(k_\perp,-i\nabla_y)$ for zero energy solutions at $\bk_\parallel=0$ and projects $H_{\parallel}(\bk_\parallel)$ into the subspace of these solutions. At $\bk_\parallel=0$ we find two solutions from which we construct second quantized operators $\gamma_{\bk_\parallel 1} $ and  $\gamma_{\bk_\parallel 2}$ satisfying the Majorana condition $\gamma^\dagger_{\bk_\parallel 1,2} = \gamma_{-\bk_\parallel 1,2}$. Projecting $H_{\parallel}(\bk_\parallel)$ into the subspace of $\gamma_{\bk_\parallel} = (\gamma_{\bk_\parallel 1} ,\gamma_{\bk_\parallel 2} )^T$ we obtain
\be
 \mathcal H_{\parallel} = -\frac{\Delta_0}{2\kf}\sum_{\bk_\parallel}  \gamma_{-\bk_\parallel}^T \left( \lambda_a k_x I_2 + \lambda_b k_z \tilde s_z  \right) \gamma_{\bk_\parallel}, \label{eq:HsurfaceC6}
\ee
where $\tilde s_z = \pm 1$ labels the surface Majorana degree of freedom and $I_2$ is a $2\times 2$ identity matrix. The profile of zero energy states is simply obtained as $|\lambda_a k_x| = |\lambda_b k_z|$, in agreement with Eq. \eqref{eq:arcsC6} and Fig. \ref{fig:arcs}.

Next, we consider the case of the $C_3$-symmetric superconductor with gap function \eqref{eq:deltaC3}. As discussed in the Section ``Symmetry analysis of quasiparticle gap structures'', the gap structure consists of eight single Majorana bulk nodes shown in Fig. \ref{fig:nodes} (and Fig. \ref{fig:schematicarcs} including projections onto surface momentum space). An analytical expression for the zero-energy mode profile analogous to Eq. \eqref{eq:arcsC6} cannot be obtained in this case, and we solve the characteristic polynomial numerically. The result is shown in Figs. \ref{fig:arcs}(c)-(d).

Similarly to the case of $C_6$ superconductors, the Hamiltonian of the surface Majorana arcs can be constructed in the vicinity of $\bk_{\parallel} = 0$ within the Andreev approximation. Again, we obtain two surface state Majorana operators $\gamma_{\bk_\parallel 1,2}$. The Hamiltonian of the Majorana arcs reads as
\be
\mathcal H_{\parallel} = \frac{\Delta_0}{2\kf}\sum_{\bk}  \gamma_{-\bk}^T \left( A_x k_x I_2 + \sqrt{\lambda_c^2 k_x^2 + A_z^2k_z^2}\tilde s_z  \right) \gamma_{\bk}, \label{eq:HparallelEu}
\ee
where the coefficients $A_x$ and $A_z$ are given by $A_x = (\lambda_c^2-\lambda_a^2)/\sqrt{\lambda_a^2+\lambda_c^2}$, $A_z = \lambda_a^4\lambda_b^2/(\lambda_a^2+\lambda_c^2)^2.$ The zero-energy states of the Majorana arcs in the vicinity of $\bk_{\parallel} = 0$ are then given by the equation $k_x^2(\lambda_a^2-3\lambda_c^2)(\lambda_a^2+\lambda_c^2) = k_z^2\lambda_a^2\lambda_b^2$ and exist only provided $\lambda_a^2>3\lambda_c^2$. Only in this range of parameters Hamiltonian (\ref{eq:HparallelEu}) is meaningful (otherwise, there is no low-energy excitations near $\bk_{\parallel} = 0$). This result is in total correspondence with the exact numerical solution shown in Figs. \ref{fig:arcs}(c)-(d).

To conclude, we consider a surface boundary in the $xy$ plane. In this geometry, the bulk $C=\pm 2$ nodes of the $C_6$ superconductor  project to a single surface momentum $k_x = k_y = 0$, such that no well-defined Majorana arcs exist in this case. In contrast, the projections of the six off-axis $C=\pm 1$ nodes of the $C_3$ superconductor do not coincide and are connected by Majorana arc states, as shown in Fig. \ref{fig:arcs}(b). As is clear from Fig. \ref{fig:arcs}(b), threefold rotational symmetry is preserved.

The above analysis shows that spin-orbit coupled chiral superconductors are topological nodal systems with characteristic surface arc states. Chiral superconductors hosting Majorana fermions in the bulk have spin-non-degenerate Majorana arcs on surface boundaries, unlike complex fermions in Weyl semimetals or spin-degenerate boundary states in superfluid $^3\text{He}$-$A$.

\subsection*{Candidate materials \label{sec:materials}}

The purpose of this section is to connect the general theory of spin-orbit coupled odd-parity chiral superconductors presented in the Section ``Symmetry analysis of quasiparticle gap structures'' to reported experimental evidence for chiral and nodal pairing in certain known superconductors. Since chiral non-unitary pairing critically relies on spin-orbit coupling, we focus the search for candidate materials hosting Majorana fermions on materials with spin-orbit coupling.

Heavy fermion materials are typically strongly spin-orbit coupled, and the vast majority of known heavy fermion superconductors are believed to have unconventional pairing symmetry. Of particular interest to the present study are two heavy fermion compounds with filled skutterudite structure: PrOs$_4$Sb$_{12}$ \cite{bauer02,aoki02} and PrPt$_4$Ge$_{12}$ \cite{gumeniuk08}. For both materials signatures consistent with point nodes have been observed, although the determination of the pairing state is not yet definitive and the Majorana nature of nodal quasiparticles remains to be tested.

The skudderudite superconductor PrOs$_4$Sb$_{12}$ has tetrahedral crystal structure with point group $T_h$. Thermal transport measurements are indicative of a superconducting phase with point nodes \cite{izawa03}. The presence of point nodes has been further corroborated by the temperature dependence of the specific heat \cite{bauer02,frederick05}, the penetration depth and NMR spin relaxation rate \cite{chia03}, and especially Sb-NQR \cite{katayama07} finding spin relaxation rate proportional to $T^5$ at temperatures considerably below $T_c$.
In addition, $\mu$SR measurements have been interpreted as supporting time-reversal symmetry breaking in the superconducting state \cite{aoki03}, and Knight shift measurements are suggestive of triplet pairing \cite{higemoto07}. Very recent Kerr angle measurements provide even more support for time-reversal symmetry breaking in the superconducting state \cite{kapitulnik16}.

A number of theoretical studies have proposed unconventional pairing symmetries as possible descriptions of the superconducting phases in PrOs$_4$Sb$_{12}$ \cite{sergienko04,abualrub07}. It has been argued that, assuming broken time-reversal symmetry and the existence of point nodes, in order to best fit experiments, the phenomenological order parameter should be of three-component $T_u$ symmetry \cite{abualrub07}. The time-reversal symmetry broken phase then corresponds to the chiral combination (i.e., phase difference $e^{\pm i \pi /2}$) of two components with different amplitude. Within this framework, the resulting chiral superconductor is a non-unitary pairing state with twofold rotational symmetry and non-degenerate point nodes not pinned at a twofold axis. Such quasiparticle spectrum can be captured by the pairing gap function given by Eq. \eqref{eq:deltaC2}. As a result, the heavy fermion superconductor PrOs$_4$Sb$_{12}$ is a promising candidate for realizing the off-axis gapless Majorana fermions.

Another member of the family of filled skutterudites with the same crystal structure as PrOs$_4$Sb$_{12}$ is the material PrPt$_4$Ge$_{12}$. Superconductivity has been observed in PrPt$_4$Ge$_{12}$ \cite{gumeniuk08}, and penetration depth in combination with specific heat measurements have provided evidence of point-like nodes \cite{maisuradze09,huang14}. Furthermore, a subsequent $\mu$SR study has reported a spontaneous magnetization below $T_c$, which is suggesting time-reversal symmetry breaking \cite{maisuradze10}. These results point towards PrPt$_4$Ge$_{12}$ as a second candidate to host gapless Majorana fermions. An NQR spin relaxation study has, however, provided support for a weakly coupled BCS superconductor with anisotropic $s$-wave pairing gap with point nodes \cite{kanetake10}. In addition, it should be noted that some experimental studies have found evidence for two-band superconductivity in PrPt$_4$Ge$_{12}$ \cite{chandra12,nakamura12,zhang13}, similar to the case of PrOs$_4$Sb$_{12}$ \cite{seyfarth05,seyfarth06,shu09}. As discussed when we considered the experimental detection of Majorana fermions, this may still be consistent with non-unitary superconductivity.

\section*{Discussion \label{sec:conclusion}}

A central pillar of this paper is the fact that time-reversal breaking pairing in crystals with strong spin-orbit coupling is generically non-unitary, leading to a (spin) non-degenerate quasiparticle gaps structure. In spin-orbit coupling systems, chiral pairing channels are labeled by the total angular momentum $J$ of the Cooper pairs, and the gap function is a general linear superposition of spherical harmonics degenerate in the pairing channel. As shown in the Section ``Symmetry analysis of quasiparticle gap structures'', the resulting gap function typically is non-unitary.

A consequence of the non-unitary nature of the pairing is the presence of non-degenerate point nodes, which satisfy the same Majorana reality condition as in high-energy particle physics and therefore constitute Majorana fermion quasiparticles in three dimensions. Depending on the (discrete) $n$-fold rotational symmetry of the crystal, and the angular momentum $j$ of the Bloch electrons, these Majorana nodes can be single nodes with linear dispersion, or double nodes with quadratic dispersion tangential to the Fermi surface. In addition, symmetry may pin the nodes to the rotation axis, in which case we call them on-axis nodes. When the rotational symmetry of the crystal is low, the on-axis nodes can be split and off-axis nodes, i.e., nodes at generic non-rotation invariant Fermi surface momenta, can appear. The splitting of Majorana nodes and the appearance of off-axis nodes is determined by the conservation of the topological monopole charge associated with point nodes. As such, it is an analog of the trigonal warping of 2D Dirac fermions in bilayer graphene \cite{falko}. Here, we find both trigonal and orthorhombic warping of Majorana nodes.

We note that our symmetry-based approach is general and complete in the sense of treating the general case of Bloch electrons with angular momentum $j$ forming Cooper pairs with total angular momentum $J$ in $C_n$-symmetric crystals. In particular, our analysis applies to superconductors where the pairing is not between spin-$\frac{1}{2}$ electrons, but more generally between spin-$j$ electrons such as $j=\frac{3}{2}$ as in half-Heusler compounds \cite{brydon16,kim16}, or $j=\frac{5}{2}$ \cite{nomoto16}.

Experimental manifestations of non-unitary pairing, and consequently of Majorana nodal fermions, can be looked for by means of probes  sensitive to the difference in spin $\up$ and $\down$ gap structure. For instance, the density of states clearly reflects the non-degeneracy of the quasiparticle spectrum, giving rise to two-gap features reminiscent of multi-band superconductors. Perhaps most prominently, we demonstrated that the NMR spin-relaxation rate in non-unitary superconductors with spin-selective low-energy quasiparticle excitations shows a marked anisotropic dependence on the polarization of the nuclear spin. We find that for a nuclear spin initially polarized along $z$ the relaxation rate is significantly suppressed, which serves as a discriminating feature of non-unitary superconductivity.

Chiral superconductors hosting Majorana fermions belong to the class of topological nodal superconductors. Topological nodal superconductors are analogs of topological semimetals called Weyl semimetals. Similarly to the latter, topological nodal superconductors have special surface states: Majorana arcs in momentum space connecting projections of bulk nodes. Whereas in unitary superconductors the Majorana arcs come in (spin-)degenerate pairs (effectively forming Fermi arcs), non-unitary superconductors have non-degenerate Majorana arcs at surface boundaries, as demonstrated in the previous section.

Most superconductors extensively studied in the literature are examples of unitary pairing states. Comparatively, non-unitary superconductors have received less attention. As discussed in the first part of the results, from a conceptual standpoint, relying on symmetry arguments, non-unitary pairing is natural and potentially widespread in spin-orbit coupled systems. We propose the heavy fermion superconductor PrOs$_4$Sb$_{12}$ as a promising candidate of such non-unitary pairing, and consequently as a realization of Majorana fermions in three dimensions.

\section*{Materials and Methods}

\subsection*{Details on Symmetries and BCS-BdG mean-field theory \label{app:bcs}}

This section introduces BCS-BdG mean field theory and the implementation of symmetries from a more formal perspective. In crystals with strong spin-orbit coupling and in the presence of both time-reversal ($\Theta$) and parity ($P$) symmetries, all electronic bands remain two-fold degenerate and the bands are labeled by an effective pseudospin $\alpha = \up,\down$. 
In the presence of $\trs$, $P$, and crystal rotation symmetry $C_n$ symmetry, one can choose a a basis such that the electron $c_{\bk \alpha} $ transform under $\trs$ and $P$ as
\begin{gather}
\trs c_{\bk \alpha} \trs^{-1}  = (is_y)_{\alpha\beta}c_{-\bk \beta}, \qquad P c_{\bk \alpha} P^{-1} =  c_{-\bk \alpha}, \label{eqapp:trsinv}
\end{gather}
respectively, and under $n$-fold rotation as
\begin{gather}
C_n c_{\bk \alpha} C_n^{-1} = (U^\dagger_n)_{\alpha\beta} \, c_{ \bk^* \beta}, \quad U_n=  \begin{pmatrix} e^{-i\theta j} & 0 \\ 0  & e^{ i\theta j}\end{pmatrix}, \label{eqapp:C_n}
\end{gather}
where here $\bk^*=C_n\bk$. Furthermore, $\theta = 2\pi/n$ is the angle of rotation and $\pm j$ is the total angular momentum of the Bloch electrons $c_{\bk \alpha} $. Note that as a result $\alpha = \up,\down$ labels the general angular momentum $\pm j$ states.

The normal state Hamiltonian is given by $\mathcal{H}_0 =   \sum_{\bk} c^\dagger_{\bk } H_0 (\bk) c_{\bk } $ with $H_0 (\bk)  =  \xi_\bk \delta_{\alpha\beta}$. 
The energy relative to the chemical potential, given by $ \xi_\bk= \varepsilon_\bk - \mu$, is a scalar function of momentum composed of terms invariant under the crystal symmetry group. The invariance of the normal state Hamiltonian under $n$-fold rotation $C_n$ is explicitly expressed as $U_{n} H_0 (\bk) U^\dagger_{n}  = H_0 (C_n\bk)$.

In a BCS-BdG mean-field theory formulation, the pairing Hamiltonian $\mathcal{H}_\Delta$ is expressed as
\begin{gather}
\mathcal{H}_\Delta =   \sum_\bk \;  ( i \Delta_\bk s_y )_{\alpha\beta}c^\dagger_{\bk\alpha}c^\dagger_{-\bk\beta} +\text{H.c.}, \label{eqapp:pairing}
\end{gather}
where the pairing matrix $\Delta_\bk$ is the momentum dependent gap function and $s_{x,y,z}$ are Pauli spin matrices acting on $\alpha = \up,\down$. We define the Nambu spinor $\Phi_\bk $ in the canonical basis in the following way
\begin{gather}
\Phi_\bk  = \begin{pmatrix} c_{\bk\alpha} \\ \epsilon_{\alpha\beta} c^\dagger_{-\bk\beta}
\end{pmatrix}. \label{eqapp:nambu}
\end{gather}
In terms of the Nambu spinor the mean-field theory BdG Hamiltonian takes the form $\mathcal{H}_{\text{BdG}}  = (1/2) \sum_\bk   \Phi^\dagger_\bk H_{\text{BdG}}(\bk) \Phi_\bk$ with
\begin{gather}
 H_{\text{BdG}}(\bk) =\begin{pmatrix} \xi_\bk & \Delta_\bk \\
   \Delta^\dagger_\bk   &  -\xi_\bk  \end{pmatrix} . \label{eqapp:bdg}
\end{gather}
The pairing Hamiltonian transforms as $\trs \mathcal{H}_\Delta  \trs^{-1}$ under time-reversal, which implies for the gap function: $\Delta_{\bk} \rightarrow (is_y)\Delta^*_{- \bk }( -is_y)$. Chiral superconductors break time-reversal symmetry and one has $(is_y)\Delta^*_{-\bk } ( -is_y) \neq \Delta_{\bk}$. Similarly, odd-parity pairing defined as $\inv \mathcal{H}_\Delta \inv^{-1} = -\mathcal{H}_\Delta$ implies $\Delta_{-\bk}=-\Delta_\bk$ for the pairing gap function. As a result of the latter, the BdG Hamiltonian possesses an effective $Z_2$ symmetry given by $\tau_z P$ such that $\tau_z P H_{\text{BdG}}(\bk)  (\tau_z P)^{-1} = H_{\text{BdG}}(-\bk)$.
In addition to these symmetries, the BdG Hamiltonian manifestly obeys a particle-hole symmetry $\phs \mathcal{H}_{\text{BdG}} \phs^{-1} = -\mathcal{H}_{\text{BdG}} $, which implies for $H_{\text{BdG}}(\bk)$: $\tau_x H_{\text{BdG}}(\bk)\tau_x =  -H^*_{\text{BdG}}(-\bk)$.

\subsection*{Details on chiral non-unitary superconductors with angular momentum $J$ \label{app:chiral}}

This paper studies odd-parity chiral superconductors in which the Cooper pairs have total angular momentum $J= L + S$. Due to spin-orbit coupling only total angular momentum is a good quantum number. The pairing gap function of angular momentum $J$ chiral superconductors is defined through the $n$-fold rotation $C_n$, which we assume to be a rotation about the $z$ axis. Concretely, the pairing gap function satisfies
\begin{gather}
U^\dagger_{n} \Delta_{C_n\bk} U_{n}  = e^{i \theta J}\Delta_\bk, \quad \theta = \frac{2\pi}{n} \label{eqapp:chiralsc}
\end{gather}
In a crystal with discrete $C_n$ rotation symmetry angular momentum is only defined mod $n$, which is manifest in Eq.~\eqref{eqapp:chiralsc}. As a consequence of Eq. \eqref{eqapp:chiralsc}, total angular momentum $J = L +S$ labels the different pairing channels of the chiral superconductors. Specifically, the gap function takes the general form of Eq. \eqref{eq:delta_k}, and all functions $F^J_t(\bk)$ with combinations $(L,S)$ such that $L+S=J$ are allowed to mix in with coefficient $\lambda_t$. 



The superconducting gap function of Eq. \eqref{eq:delta_k} can be explicitly expanded in the spin matrices $s_{x,y,z}$ as $\Delta_\bk  = \bd(\bk)\cdot \boldsymbol{s}$, where $\bd(\bk)$ is momentum dependent vector. One then finds for $\Delta^\dagger_\bk \Delta_\bk$
\begin{gather}
\Delta^\dagger_\bk \Delta_\bk = |\bd|^2 I_2 + i \bd^*\times \bd \cdot \boldsymbol{s}. \label{eqapp:nonunitary}
\end{gather}
When $\bd^*\times \bd  = 0$ the pairing is said to be unitary, and the quasiparticle spectrum of unitary pairing states is manifestly twofold (spin-)degenerate. In contrast, when $\bd^*\times \bd  \neq 0$ the pairing is said to be non-unitary and the quasiparticle energies are given by $E_{\bk \pm} = ( \xi_\bk^2 +|\bd |^2 \pm |\bd^*\times \bd |)^{1/2}$. Consequently, non-unitary pairing states are characterized by a non-degenerate quasiparticle energy dispersion, with different gap structures for the spin $\up$ and spin $\down$ electrons.

As a consequence of spin-orbit coupling and the lack of spin-rotation symmetry, a generic gap function given by Eq. \eqref{eq:delta_k} corresponds to \emph{non-unitary} pairing states. This is easily seen with the help of Table \ref{tab:gapfunctions}.

\subsection*{Details on NMR spin relaxation calculation \label{ssec:nmr}}

This section briefly recapitulates how the NMR relaxation rate is obtained. We calculate the NMR relaxation rate using Fermi's Golden Rule, which can be shown to be equivalent to Eq. \eqref{eq:T1}. Defining the nuclear spin coherent state $| \bS \rangle$ for a spin initially polarized along $\bS = (\sin \theta \cos \varphi, \sin \theta \sin \varphi, \cos \theta)$ as $| \bS \rangle = [\cos (\theta/2) , e^{i\varphi} \sin (\theta / 2)]^T$ (we consider nuclear spin 1/2 for simplicity), Fermi's golden rule for the NMR relaxation takes the form
\begin{multline}
\frac1{T_1} = 2\pi\sum_{\bk \bk' s s'}|\langle -\bS , a_{\bk s}   |\mathcal H_{\text{hf}}| \bS ,a_{\bk' s' }\rangle |^2 \, \times \\
f_{\bk' s'}(1-f_{\bk s}) \delta(E_{s'\bk} -E_{s\bk'}), \label{eqapp:FGR}
\end{multline}
where $a_{\bk s}$ are Bogoliubov quasiparticles, $f_{\bk}$ is Fermi-Dirac distribution function, and $E_{1,2}(\bk)$ are eigenenergies of BdG Hamiltonian.

\subsection*{Details on Majorana arcs in the Andreev approximation  \label{ssec:majoranaarcs}}

This section explains how the Majorana arcs are calculated within the Andreev approximation. To derive the Majorana arc surface states within the semiclassical or Andreev approximation (following Ref. \onlinecite{park15}), we solve the BdG Hamiltonian (\ref{eqapp:bdg}) in the presence of a spatially dependent pairing potential $\Delta_{\bk}(\br)$. Specifically, we assume that $\Delta_\bk(\br)$ is given by $\Delta_\bk(\br) = \Delta_\bk \Theta(y)$, i.e., a surface boundary in the $xz$ plane. Substituting $\bk \rightarrow -i \makebf{\nabla}$, the first-quantized BdG equation reads as
\begin{gather}
H_{\text{BdG}}(-i \makebf{\nabla})  \Psi(\br)   = E  \Psi(\br) . \label{eqapp:hamsurface}
\end{gather}
Here $\Psi(\br)$ is a (first-quantized) spinor wave function which we further decompose into $\Psi(\br)   = \psi_{\bk_\parallel,\pm}(\br ) \Psi_0$, where $ \Psi_0$ is a spinor, $\bk_\parallel = (k_x,k_z)$ is the momentum parallel to the boundary surface, which is a good quantum number, and the functions $\psi_{\bk_\parallel,\pm}(\br ) $ take the general form
\begin{gather}
\psi_{\bk_\parallel,\pm}(\br )   =  \frac{1}{N} e^{i \bk_\parallel\cdot\br_\parallel} e^{\pm i k_\perp y} \chi(y)
\end{gather}
Here $N$ is a normalization constant, the parallel coordinates are given by $\br_{\parallel} = (x,z)$, and $k_\perp = (k^2_F - k^2_\parallel)^{1/2}$ is defined as the semiclassical momentum perpendicular to the boundary surface. $\chi(y)$ is a scalar function. We will demand that the wave functions satisfy the Dirichlet boundary condition at $y=0$ and $y \to \infty$, i.e., $\left.\Psi(\br)\right\vert_{y=0}=0$ and $\left.\Psi(\br)\right\vert_{y\to \infty}=0$. The function $\chi(y)$ will always be such that the latter is satisfied, and to satisfy the former we take superpositions of incident and reflected waves.

In the semiclassical approximation, defined by the condition $k_\perp \gg k_F (\Delta_0/\varepsilon_F)$, the BdG equation \eqref{eqapp:hamsurface} takes the form of the Andreev equation for $\chi(y)$, given by
\begin{gather}
E \chi(y) \Psi_0 = \left[ H_0(\bk_\parallel) \pm H_\perp(k_\perp ,-i\nabla_y) \right] \chi(y) \Psi_0. \label{eqapp:hamss}
\end{gather}
Here the Hamiltonian $H_0(\bk_\parallel)$ depends only on the momenta $\bk_\parallel$ and $H_\perp(k_\perp ,-i\nabla_y)$ is a function of $\hat p _y = -i\nabla_y$ and the semiclassical momentum $k_\perp$. Our strategy will be to solve Eq. \eqref{eqapp:hamss} for the case $k_\parallel = 0$ and $E=0$, and then obtain the effective Hamiltonian of the Majorana arcs in the vicinity of $k_\parallel = 0$ by projecting $H_0(\bk_\parallel)$ into the space of zero energy solutions of $H_\perp(k_\perp ,-i\nabla_y)$.

\bibliographystyle{apsrev}

\begin{thebibliography}{10}

\bibitem{kallin} C. Kallin, Rep. Prog. Phys. {\bf 75}, 042501 (2012).

\bibitem{maeno} A. Mackenzie and Y. Maeno, Rev. Mod. Phys. {\bf 75}, 657 (2003).

\bibitem{leggett75} A. J. Leggett, Rev. Mod. Phys. {\bf 47}, 331 (1975).

\bibitem{balents} L. Balents, Physics {\bf 4}, 36 (2011).

\bibitem{volovik-weyl} G. E. Volovik, JETP Letters, {\bf 103}, 140 (2016).


\bibitem{venderbos15} J. W. F. Venderbos, V. Kozii, and L. Fu, arXiv:1512.04554 (2015).


\bibitem{wilczek} F. Wilczek, Nat. Phys. {\bf 5}, 614 (2009).

\bibitem{beenakker13} C. W. J. Beenakker, Annu. Rev. Condens. Matter Phys. {\bf 4}, 113 (2013).

\bibitem{sigristueda} M. Sigrist and K. Ueda, Rev. Mod. Phys. {\bf 63}, 239 (1991).



\bibitem{ohmi96} Ohmi, T., and K. Machida, 1996a, J. Phys. Soc. Jpn. {\bf 65}, 3456 (1996); {\it Ibid},  J. Phys. Soc. Jpn. 65, 4018 (1996).

\bibitem{hou98} H. Tou, Y. Kitaoka, K. Ishida, K. Asayama, N. Kimura, Y. Onuki, E. Yamamoto, Y. Haga, and K. Maezawa
Phys. Rev. Lett. {\bf 80}, 3129 (1998).

\bibitem{sauls94} J. A. Sauls, Advances in Physics {\bf 43}, 113 (1994).

\bibitem{joynt02} R. Joynt and L. Taillefer, Rev. Mod. Phys. {\bf 74}, 235 (2002).


\bibitem{sigrist96}  M. Sigrist and M. E. Zhitomirsky, J. Phys. Soc. Jpn. {\bf 65}, 3452 (1996).

\bibitem{machida96} K. Machida, M. Ozaki, and T. Ohmi, J. Phys. Soc. Jpn. {\bf 65}, 3720 (1996).



\bibitem{weng16} Z. F. Weng, J. L. Zhang, M. Smidman, T. Shang, J. Quintanilla, J. F. Annett, M. Nicklas, G. M. Pang, L. Jiao, W. B. Jiang, Y. Chen, F. Steglich, and H. Q. Yuan, Phys. Rev. Lett. {\bf 117}, 027001 (2016).

\bibitem{quintanilla10} J. Quintanilla, A. D. Hillier, J. F. Annett, and R. Cywinski, Phys. Rev. B {\bf 82}, 174511 (2010)



\bibitem{ambegaokar73} V. Ambegaokar and N. D. Mermin, Phys. Rev. Lett. {\bf 30}, 81 (1973).

\bibitem{yip93} S. Yip and A. Garg, Phys. Rev. B { \bf 48}, 3304 (1993).


\bibitem{brydon16} P. M. R. Brydon, L. Wang, M. Weinert, and D. F. Agterberg, Phys. Rev. Lett. {\bf 116}, 177001 (2016).

\bibitem{kim16} H. Kim, K. Wang, Y. Nakajima, R. Hu, S. Ziemak, P. Syers, L. Wang, H. Hodovanets, J. D. Denlinger, P. M. R. Brydon, D. F. Agterberg, M. A. Tanatar, R. Prozorov, J. Paglione, arXiv:1603.03375 (2016).


\bibitem{nomoto16} T. Nomoto and H. Ikeda, arXiv:1607.02708 (2016).

\bibitem{volovikbook} G. E. Volovik, The Universe in a Helium Droplet, Oxford: Oxford University Press (2003).

\bibitem{mizushima16} T. Mizushima, Y. Tsutsumi, T. Kawakami, M. Sato, M. Ichioka, and K. Machida, J. Phys. Soc. Jpn. { \bf 85}, 022001 (2016).





\bibitem{meng12} T. Meng and L. Balents, Phys. Rev. B {\bf 86}, 054504 (2012).

\bibitem{sau12} J. D. Sau, S. Tewari, Phys. Rev. B {\bf 86}, 104509 (2012).

\bibitem{fischer14}  M. H. Fischer, T. Neupert, C. Platt, A. P. Schnyder, W. Hanke, J. Goryo, R. Thomale, and M. Sigrist, Phys. Rev. B {\bf 89}, 020509(R) (2014).

\bibitem{wang12} F. Wang and D. H. Lee, Phys. Rev. B {\bf 86} 094512 (2012).

\bibitem{sato06} M. Sato, Phys. Rev. B {\bf 73} 214502 (2006).

\bibitem{beri10} B. B\'eri, Phys. Rev. B {\bf 81} 134515 (2010).

\bibitem{schnyder11} A. P. Schnyder and S. Ryu, Phys. Rev. B {\bf 84} 060504 (2011). 

\bibitem{brydon11} P. M. R. Brydon, A. P. Schnyder and C. Timm, Phys. Rev. B {\bf 84} 020501 (2011). 

\bibitem{schnyder12} A. P. Schnyder,P. M. R. Brydon and C. Timm, Phys. Rev. B {\bf 85} 024522 (2012). 

\bibitem{sato11} M. Sato, Y. Tanaka, K. Yada, and T. Yokoyama, Phys. Rev. B {\bf 83} 224511 (2011). 

\bibitem{yada11} K. Yada, M. Sato, Y. Tanaka, and T. Yokoyama, Phys. Rev. B {\bf 83} 064505 (2011). 



\bibitem{tanaka10}  Y. Tanaka, Y. Mizuno, T. Yokoyama, K. Yada, and M. Sato, Phys. Rev. Lett. {\bf 105} 097002 (2010). 

\bibitem{sato10} M. Sato and S. Fujimoto, Phys. Rev. Lett. {\bf 105} 217001 (2010). 



\bibitem{cho12} G. Y. Cho, J. H. Bardarson, Y.-M. Lu, and J. E. Moore, Phys. Rev. B {\bf 86}, 214514 (2012).

\bibitem{shivamoggi14} V. Shivamoggi and M. J. Gilbert, Phys. Rev. B {\bf 88}, 134504 (2013).

\bibitem{wei14} H. Wei, S.-P. Chao, and V. Aji, Phys. Rev. B {\bf 89}, 014506 (2014).

\bibitem{zhang14} S. A. Yang, H. Pan, and F. Zhang, Phys. Rev. Lett. {\bf 113}, 046401 (2014).

\bibitem{kobayashi15} S. Kobayashi and M. Sato, Phys. Rev. Lett. {\bf 115}, 187001 (2015).

\bibitem{lu15b} B. Lu, K. Yada, M. Sato, and Y. Tanaka, Phys. Rev. Lett. {\bf 114} , 096804 (2015).

\bibitem{bednik15} G. Bednik, A. A. Zyuzin, and A. A. Burkov, Phys. Rev. B {\bf 92}, 035153 (2015).

\bibitem{li15} Yi Li and F. D. M. Haldane, arXiv:1510.01730 (2015).





\bibitem{wan11} X. Wan, A. M. Turner, A. Vishwanath, and S. Y. Savrasov, Phys. Rev. B {\bf 83}, 205101  (2011).

 \bibitem{turner13} A. M. Turner, A. Vishwanath, Chapter 10, {\it Contemporary Concepts of Condensed Matter Science. Volume 6: Topological Insulators}, M. Franz and L. Molenkamp (Editors), (Elsevier, 2013).

\bibitem{fang12} C. Fang, M. J. Gilbert, X. Dai, and B. A. Bernevig, Phys. Rev. Lett. {\bf 108}, 266802 (2012).



\bibitem{matsuda06} Y. Matsuda, K. Izawa, and I. Vekhter, J. Phys.: Condens. Matter {\bf 18}, R705 (2006).


\bibitem{hebel59} L. C. Hebel and C. P. Slichter, Phys. Rev. {\bf 113}, 1504 (1959).



\bibitem{moriya} T. Moriya, J. Phys. Soc. Jpn. {\bf 18}, 516 (1963).

\bibitem{chung09} S. B. Chung and S.-C. Zhang, Phys. Rev. Lett. {\bf 103} 235301 (2009).

\bibitem{nagato09} Y. Nagato, S. Higashitani, and K. Nagai, J. Phys. Soc. Jpn. {\bf 78}, 123603 (2009).


\bibitem{matsuura13} S. Matsuura, P.-Y. Chang, A. P. Schnyder, and S. Ryu, New J. Phys. 15, 065001 (2013).


\bibitem{tsutsumi11} Y. Tsutsumi, M. Ichioka, and K. Machida, Phys. Rev. B {\bf 83}, 094510 (2011).

\bibitem{silaev12} M. A. Silaev and G. E. Volovik, Phys. Rev. B {\bf 86}, 214511 (2012).


\bibitem{goswami15} P. Goswami and A. H. Nevidomskyy, Phys. Rev. B {\bf 92}, 214504 (2015).


\bibitem{burkov11} A. A. Burkov and L. Balents, Phys. Rev. Lett. {\bf 107}, 127205 (2011).

\bibitem{burkov11b} A. A. Burkov, M. D. Hook, and L. Balents, Phys. Rev. B {\bf 84}, 235126
(2011).

\bibitem{hosur12} P. Hosur, Phys. Rev. B {\bf 86}, 195102 (2012).

\bibitem{weng15}  H. Weng, C. Fang, Z. Fang, B. A. Bernevig, and X. Dai, Phys. Rev. X {\bf 5}, 011029 (2015).




\bibitem{huang15}  S.-M. Huang, S.-Y. Xu, I. Belopolski, C.-C. Lee, G. Chang, B. Wang, N. Alidoust, G. Bian, M. Neupane,  C. Zhang, S. Jia, A. Bansil, H. Lin, M. Z. Hasan, Nat. Comm. {\bf 6}, 7373 (2015).

\bibitem{xu15} S.-Y. Xu, I. Belopolski, N. Alidoust, M. Neupane, G. Bian, C. Zhang, R. Sankar, G. Chang, Z. Yuan, C.-C. Lee, S.-M. Huang, H. Zheng, J. Ma, D. S. Sanchez, B. Wang, A. Bansil, F. Chou, P. P. Shibayev, H. Lin, S. Jia, M. Z. Hasan, Science {\bf 349}, 613 (2015).

\bibitem{xu15b} S.-Y. Xu, N. Alidoust, I. Belopolski, Z. Yuan, G. Bian, T.-R. Chang, H. Zheng, V. N. Strocov, D. S. Sanchez, G. Chang, C. Zhang, D. Mou, Y. Wu, L. Huang, C.-C. Lee, S.-M. Huang, B Wang, A. Bansil, H.-T. Jeng, T. Neupert, A. Kaminski, H. Lin, S. Jia, M. Z. Hasan, Nat. Phys. {\bf 11}, 748 (2015).

\bibitem{lv15} B. Q. Lv, H. M. Weng, B. B. Fu, X. P. Wang, H. Miao, J. Ma, P. Richard, X. C. Huang, L. X. Zhao, G. F. Chen, Z. Fang, X. Dai, T. Qian, H. Ding, Phys. Rev. X {\bf 5}, 031013 (2015).

\bibitem{lv15b} B. Q. Lv, N. Xu, H. M. Weng, J. Z. Ma, P. Richard, X. C. Huang, L. X. Zhao, G. F. Chen, C. E. Matt, F. Bisti, V. N. Strocov, J. Mesot, Z. Fang, X. Dai, T. Qian, M. Shi, H. Ding, Nat. Phys. {\bf 11}, 724 (2015).

\bibitem{yang15} L. X. Yang, Z. K. Liu, Y. Sun, H. Peng, H. F. Yang, T. Zhang, B. Zhou, Y. Zhang, Y. F. Guo, M. Rahn, D. Prabhakaran, Z. Hussain, S.-K. Mo, C. Felser, B. Yan, Y. L. Chen, Nat. Phys. {\bf 11}, 728 (2015).

\bibitem{bernevig15} B. A. Bernevig, Nat. Phys. {\bf 11}, 698 (2015).

\bibitem{lu13} L. Lu, L. Fu, J. D. Joannopoulos, and M. Soljacic, Nat. Photon.  {\bf 7}, 294 (2013).

\bibitem{lu14} L. Lu, J. D. Joannopoulos, and M. Soljacic, Nat. Photon. {\bf 8}, 821 (2014).

\bibitem{lu15} L. Lu, Z. Wang, D. Ye, L. Ran, L. Fu, J. D. Joannopoulos, and M. Soljacic, Science {\bf 349}, 622 (2015).


\bibitem{young12}  S. M. Young, S. Zaheer, J. C. Y. Teo, C. L. Kane, E. J. Mele, and A. M. Rappe, Phys. Rev. Lett. {\bf 108}, 140405 (2012).


\bibitem{xu11} G. Xu, H. Weng, Z. Wang, X. Dai, and Z. Fang, Phys. Rev. Lett. {\bf 107}, 186806 (2011).

\bibitem{yang11} K.-Y. Yang, Y.-M. Lu, and Y. Ran, Phys. Rev. B {\bf 84}, 075129 (2011).

\bibitem{krempa12} W. Witczak-Krempa and Y. B. Kim, Phys. Rev. B {\bf 85}, 045124 (2012).

\bibitem{kim13} H.-J. Kim, K.-S. Kim, J.-F. Wang, M. Sasaki, N. Satoh, A. Ohnishi, M. Kitaura, M. Yang, and L. Li, Phys. Rev. Lett. {\bf 111}, 246603 (2013).

\bibitem{bulmash14} D. Bulmash, C.-X. Liu, and X.-L. Qi, Phys. Rev. B {\bf 89}, 081106(R) (2014).

\bibitem{dubcek15} T. Dubcek, C. J. Kennedy, L. Lu, W. Ketterle, M. Soljacic, and H. Buljan, Phys. Rev. Lett. {\bf 114}, 225301 (2015).

\bibitem{borisenko15} S. Borisenko, D. Evtushinsky, Q. Gibson, A. Yaresko, T. Kim, M. N. Ali, B. B\"uchner, M. Hoesch, and R. J. Cava, arXiv preprint arXiv:1507.04847 (2015).

\bibitem{chang16} G. Chang, S.-Y. Xu, H. Zheng, B. Singh, C.-H. Hsu, I. Belopolski, D. S. Sanchez, G. Bian, N. Alidoust, H. Lin, M. Z. Hasan, arXiv:1603.01255 (2016).





\bibitem{park15} Y-J Park, S. B. Chung, and J. Maciejko, Phys. Rev. B {\bf 91}, 054507 (2015).



\bibitem{bauer02} E. D. Bauer, N. A. Frederick, P.-C. Ho, V. S. Zapf, and M. B. Maple, Phys. Rev. B {\bf 65}, 100506(R) (2002).

\bibitem{aoki02} Yuji Aoki, Takahiro Namiki, Shuji Ohsaki, Shanta R. Saha, Hitoshi Sugawara, and Hideyuki Sato, J. Phys. Soc. Jpn. {\bf 71}, 2098 (2002).

\bibitem{gumeniuk08} R. Gumeniuk, W. Schnelle, H. Rosner, M. Nicklas, A. Leithe-Jasper, and Yu. Grin, Phys. Rev. Lett. {\bf 100}, 017002 (2008).


\bibitem{izawa03} K. Izawa, Y. Nakajima, J. Goryo, Y. Matsuda, S. Osaki, H. Sugawara, H. Sato, P. Thalmeier, and K. Maki, Phys. Rev. Lett. {\bf 90}, 117001 (2003).

\bibitem{frederick05} N. A. Frederick, T. A. Sayles, and M. B. Maple, Phys. Rev. B {\bf 71}, 064508 (2005).

\bibitem{chia03}  E. E. M. Chia, M. B. Salamon, H. Sugawara, and H. Sato Phys. Rev. Lett. {\bf 91}, 247003 (2003).

\bibitem{katayama07}  K. Katayama, S. Kawasaki1, M. Nishiyama, H. Sugawara, D. Kikuchi, H. Sato, and G.-q,]. Zheng, J. Phys. Soc. Jpn. {\bf 76}, 023701 (2007).


\bibitem{aoki03} Y. Aoki, A. Tsuchiya, T. Kanayama, S. R. Saha, H. Sugawara, H. Sato, W. Higemoto, A. Koda, K. Ohishi, K. Nishiyama, and R. Kadono, Phys. Rev. Lett. {\bf 91}, 067003 (2003).


\bibitem{higemoto07} W. Higemoto, S. R. Saha, A. Koda, K. Ohishi, R. Kadono, Y. Aoki, H. Sugawara, and H. Sato, Phys. Rev. B {\bf 75}, 020510(R) (2007).

\bibitem{kapitulnik16} E. M. Levenson-Falk, E. R. Schemm, M. B. Maple, and A. Kapitulnik, arXiv:1609.07535 (2016).

\bibitem{sergienko04} I. A. Sergienko and S. H. Curnoe, Phys. Rev. B {\bf 70}, 144522 (2004).

\bibitem{abualrub07} T. R. Abu Alrub and S. H. Curnoe, Phys. Rev. B {\bf 76}, 054514 (2007).


\bibitem{maisuradze09}  A. Maisuradze, M. Nicklas, R. Gumeniuk, C. Baines, W. Schnelle, H. Rosner, A. Leithe-Jasper, Yu. Grin, and R. Khasanov, Phys. Rev. Lett. {\bf 103}, 147002 (2009).


\bibitem{huang14} K. Huang, L. Shu, I. K. Lum, B. D. White, M. Janoschek, D. Yazici, J. J. Hamlin, D. A. Zocco, P.-C. Ho, R. E. Baumbach, and M. B. Maple Phys. Rev. B {\bf 89}, 035145 (2014).

\bibitem{maisuradze10} A. Maisuradze, W. Schnelle, R. Khasanov, R. Gumeniuk, M. Nicklas, H. Rosner, A. Leithe-Jasper, Yu. Grin, A. Amato, and P. Thalmeier, Phys. Rev. B {\bf 82}, 024524 (2010).

\bibitem{kanetake10} F. Kanetake, H. Mukuda, Y. Kitaoka, K. Magishi, H Sugawara, K. M. Itoh, and E. E. Haller, J. Phys. Soc. Jpn. {\bf 79}, 063702 (2010).


\bibitem{chandra12} M. K. Sharath Chandra, L. S. Chattopadhyay, and S. B. Roy, Philos. Mag. {\bf 92}, 3866 (2012).

\bibitem{nakamura12} Y. Nakamura, H. Okazaki, R. Yoshida, T. Wakita, H. Takeya, K. Hirata, M. Hirai, Y. Muraoka, and T. Yokoya, Phys. Rev. B {\bf 86}, 014521 (2012).

\bibitem{zhang13} J. L. Zhang, Y. Chen, L. Jiao, R. Gumeniuk, M. Nicklas, Y. H. Chen, L. Yang, B. H. Fu, W. Schnelle, H. Rosner, A. Leithe-Jasper, Y. Grin, F. Steglich, and H. Q. Yuan, Phys. Rev. B {\bf 87}, 064502 (2013).


\bibitem{seyfarth05} G. Seyfarth, J. P. Brison, M.-A. M\'easson, J. Flouquet, K. Izawa, Y. Matsuda, H. Sugawara, and H. Sato, Phys. Rev. Lett. {\bf 95}, 107004 (2005).

\bibitem{seyfarth06} G. Seyfarth, J. P. Brison, M.-A. M\'easson, D. Braithwaite, G. Lapertot, and J. Flouquet, Phys. Rev. Lett. {\bf 97}, 236403 (2006).

\bibitem{shu09} L. Shu, D. E. MacLaughlin, W. P. Beyermann, R. H. Heffner, G. D. Morris, O. O. Bernal, F. D. Callaghan, J. E. Sonier, W. M. Yuhasz, N. A. Frederick, and M. B. Maple, Phys. Rev. B {\bf 79}, 174511 (2009).


\bibitem{falko} E. McCann and V. I. Fal'ko, Phys. Rev. Lett. {\bf 96}, 086805 (2006).



\end{thebibliography}

\subsection*{}

{\bf Acknowledgement:} We thank Guo-qing Zheng for helpful discussions.  {\bf Funding:} This work is supported by DOE Office of Basic Energy Sciences, Division of Materials Sciences and Engineering under Award DE-SC0010526 (LF and VK), and the Netherlands Organization for Scientific Research (NWO) through a Rubicon grant (JV). {\bf Author contributions:} V.K., J.W.F.V., and L.F. conceived of the problem, performed the analysis and calculations, and contributed to writing the manuscript.  {\bf Competing interests:} The authors declare that they have no competing interests. {\bf Data and materials availability:} All data needed to validate the conclusions of the paper are provided in the paper or Appendices. Additional data or information may be requested from authors.

\appendix

\section{Details on the symmetry analysis of low-energy gap structures \label{app:symmetry}}

This section provides a more detailed version of the symmetry analysis presented in the section ``Symmetry analysis of quasiparticle gap structure'' of the main text.

First, we expand the single-particle Hamiltonian $\mathcal H_0 $ near $\pm \bK$ to obtain the low-energy band dispersion in the normal state.
\be
\mathcal H_0 = \sum_{\bq, \alpha}  \xi_\bq ( c^\dagger_{ \bK+\bq \alpha} c_{ \bK+\bq \alpha} +
c^\dagger_{-\bK-\bq \alpha} c_{ -\bK -\bq \alpha}). \label{eqapp:H0expand}
\ee
Here $\xi_\bq \equiv \varepsilon_{\bK+ \bq} - \varepsilon_\bK = \varepsilon_{\bK+ \bq} - \mu$ is the single-particle energy of Bloch states with small momentum $\bq$ relative to the node at $ \bK$ and is given by
\be
\xi_\bq = v_F q_z +(q_x^2 + q_y^2)/2m \label{eq:epsq}
\ee
to lowest order in $\bq$, where $v_F$ is Fermi velocity in the $z$ direction and $m$ parametrizes the curvature of Fermi surface at $\bK$.

Next, we expand the pairing Hamiltonian (33) in small momenta $\bq$. Specifically, projecting the Pairing Hamiltonian into the subspace of the $c_{\pm\bK +\bq \alpha}$ states reads as
\begin{gather}
\mathcal H_\Delta =\sum_{\nu=\pm,\bq}  (i  \Delta_{\nu\bK+\bq} s_y)_{\alpha\beta} c^\dagger_{\nu\bK+\bq \alpha} c^\dagger_{-\nu\bK-\bq \beta}+ \text{H.c.}. \label{eqapp:pairingexpand}
\end{gather}
The pairing matrix $ \Delta_{\nu\bK+\bq} $ is then expanded in spin space and the small momenta $\bq$ as follows
\be
 \Delta_{\nu\bK+\bq}  =\frac{1}{2} \Delta_{1\nu}(\bq)s_+ + \frac{1}{2}\Delta_{2\nu}(\bq)s_- +  \Delta_{3\nu}(\bq)s_z, \label{eq:evenJ2}
\ee
where $\nu=\pm$ and $s_\pm = s_x \pm is_y $. The momentum dependent functions $ \Delta_{i\nu}(\bq)$ ($i=1,2,3$) are constrained by the odd-parity property of the gap function: $ \Delta_{\bK+\bq} = - \Delta_{-\bK-\bq} $. This leads to $ \Delta_{i + }(\bq) = -\Delta_{i - }(-\bq)$ and we need only to consider $ \Delta_{i + }(\bq) \equiv \Delta_{i }(\bq) /2$, with the gap functions $\Delta_{i }(\bq)$ as defined in Eq. (3).

Once the total angular momentum $J$ of the superconductor and the angular momentum $j$ of the Bloch states are specified, the transformation of the gap function under rotations given by Eq. (36) becomes
\begin{widetext}
\beq
U^\dagger_{n}  \Delta_{\bK+C_n\bq}  U_{n}  &  = &  \frac{1}{2}\Delta_{1}(C_n\bq) U^\dagger_{n}s_+U_n + \frac{1}{2}\Delta_{2}(C_n\bq)U^\dagger_{n}s_-U_n  +  \Delta_{3}(C_n\bq)U^\dagger_{n}s_zU_n  \nonumber \\
 & = &  e^{i\theta 2j}\frac{1}{2}\Delta_{1}(C_n\bq) s_+ + e^{-i\theta 2j}\frac{1}{2} \Delta_{2}(C_n\bq)s_-  + \Delta_{3}(C_n\bq)s_z   \nonumber \\
 & = & e^{i \theta J }\Delta_{\bK+\bq},  \label{eqapp:deltatransform}
\eeq
\end{widetext}
where $\theta = 2\pi /n$ and $U_n$ is given by Eq. (32). The three pairing functions $ \Delta_{i}(\bq)$ satisfy
\be
\Delta_{i}(C_n\bq) = e^{i \theta l_i } \Delta_{i}(\bq),
\ee
and using Eq. \eqref{eqapp:deltatransform} this implies for the orbital angular momenta $l_i$
\begin{eqnarray}
2j + l_1 &=& J \; \text{mod} \; n , \nonumber \\
-2j + l_2 &=& J \; \text{mod} \; n ,  \nonumber \\
l_3 &=& J \; \text{mod} \; n .  \label{eqapp:orbital}
\end{eqnarray}
These relations determine the structure of the pairing functions $ \Delta_{i}(\bq)$. More specifically, the pairing functions can be explicitly expanded in the tangential momenta $q_\pm = q_x \pm i q_y$ as follows (see also Ref. 48)
\be
\Delta_{i}(\bq) = \sum_{a_i,b_i} C_{a_ib_i} q^{a_i}_+q^{b_i}_- , \label{eqapp:deltaexpand}
\ee
where $C_{a_ib_i}$ are complex coefficients and $a_i,b_i$ must satisfy $a_i-b_i = l_i$ mod $n$. Since we are interested only in the terms which are of lowest order in $q_\pm$, we keep term of the form $C^{+}_i q^{a_i}_+  + C_i^-q^{b_i}_-$ in \eqref{eqapp:deltaexpand}, which gives Eq. (4) of the main text. Then $a_i,b_i$ are the smallest nonnegative integers matching orbital angular momentum $l_i$ mod $n$.

We proceed to particularize to the case when $2j = J$ mod $n$ which is the main focus of the main text (section ``Majorana nodes on rotation axis''). Since $l_1=0$ the spin $\up$ states can pair at $\bK$, and we thus focus on $\Delta_{2}(\bq) \equiv \Delta (\bq)$. With the definition of $\Psi_{\bq}$ in Eq. (6) and the normal state Hamiltonian \eqref{eqapp:H0expand}, the Hamiltonian of the low-energy $\Psi_{\bq}$ quasiparticles reads as
\begin{multline}
H(\bq)= \xi^+_\bq\tau_z  +  \xi^-_\bq \sigma_z + \frac{1}{4}( \Delta_\bq\sigma_+ - \Delta_{-\bq}\sigma_-) \tau_+ \\
+ \frac{1}{4}( \Delta^*_\bq\sigma_- - \Delta^*_{-\bq}\sigma_+) \tau_- \label{eqapp:bdglowenergy}
\end{multline}
where (as in the main text) a set of Pauli matrices $\sigma_{x,y,z}$ is introduced which act on the nodal degree of freedom given by $\pm \bK$ (i.e., $\sigma_z =\pm 1$ corresponds to $\pm \bK$). Furthermore, we defined $\xi^\pm_\bq = (\xi_\bq \pm \xi_{-\bq})/2$ and $\sigma_\pm = \sigma_x \pm i \sigma_y$ (similarly for $\tau_\pm$). In the section titled ``Symmetry analysis of quasiparticle gap structures'' of the main text we work with $\xi_\bq = v_Fq_z$ and therefore $\xi^+_\bq = 0$.

From Eq. \eqref{eqapp:orbital} one finds that $l_2 \equiv l = 4j$ mod $n$, and if in addition we define $l' = -4j$ mod $n$ (note that in this way $l,l'$ are both nonnegative integers) we can expand $\Delta_\bq$ as
\begin{gather}
\Delta_\bq = \left\{ \begin{array}{cl} C   & \quad l   = l'=0 \\ C^+ q^l_+  & \quad l   < l'  \\ C^- q^{l'}_-  & \quad l  > l' \\ C^+ q^{l}_+ + C^- q^{l'}_-  &\quad l  = l' \neq 0\end{array} \right.
\end{gather}
Here $C^\pm$ and $C$ are complex numbers. Applying the above formula to the case of $C_{6,4,3,2}$-symmetric superconductors yields Table I. All these cases are discussed in the main text, except for the $C_4$-symmetric case. The case $(n,j) = (4,\frac{1}{2})$ is an example of $l=l'= 2$, giving rise to $\Delta_\bq = C^+ q^{2}_+ + C^- q^{2}_-$. The corresponding Hamiltonian for the low-energy Majorana quasiparticles then takes the form
\begin{multline}
H(\bq)= v_Fq_z \sigma_z + \frac{1}{2}i\sigma_y(C^+ q^{2}_+ + C^- q^{2}_-) \tau_+ \\
-\frac{1}{2}i\sigma_y [ (C^+)^* q^{2}_- + (C^-)^* q^{2}_+] \tau_- \label{eqapp:C4}
\end{multline}
The monopole charge of the node at $\pm \bK$ of such Hamiltonian is given by $ \pm2 \,\text{sgn}(|C^+| - |C^-| )$, from which we conclude that in case of $C_n$ symmetry there exist double-node Majorana fermions on the rotation axis, with a sign that depends on microscopic details. A schematic representation of the sign ambiguity of the double nodes is given in Fig. 1.

In chiral superconductors a mirror symmetry $M$ of the crystal with a mirror plane that includes the rotation axis is broken. However, when the mirror operation is combined with time-reversal $\Theta$, the combined operation $\Theta M$ is a symmetry of the chiral superconductor. In this case, one can show that $C^- = (C^+)^*$ in Eq. \eqref{eqapp:C4}.

\section{Derivation of Majorana nodes}

In this section we explicitly derive the on-axis nodal Majorana quasiparticles at the $C_n$-invariant Fermi surface momenta $\pm \bK$ from the gap functions $\Delta_\bk$ defined in Eqs. (13) and (16) of the main text. To obtain the low-energy Hamiltonian of the Majorana fermions, we expand the full pairing Hamiltonian (35) in small momenta $\bq$ measured from $\pm \bK$. We define the low-energy long wavelength Nambu fields $\Psi_{\bq \alpha} $ (where $\alpha = \up,\down$)  as
\begin{gather}
\Psi_{\bq \alpha} =  \begin{pmatrix} c_{\bK +\bq \alpha}  \\ c_{-\bK +\bq \alpha}  \\ c^\dagger_{\bK - \bq \alpha} \\ c^\dagger_{-\bK -\bq \alpha}  \end{pmatrix} . \label{eqapp:majorana}
\end{gather}
With this definition both fields $\Psi_{\bq \alpha}$ satisfy the Majorana reality condition $\Psi^\dagger_{\bq \alpha} = (\tau_x \Psi_{-\bq \alpha})^T$. Clearly, $\Psi_{\bq \down}$ is the Majorana quantum field defined in Eq. (6). In terms of the fields $\Psi_{\bq \alpha}$ the Hamiltonian near $\pm \bK$ takes the general form
\begin{gather}
\mathcal{H}_{\text{BdG}}  \simeq \frac{1}{2}\sum_\bq \begin{pmatrix}  \Psi^\dagger_{\bq \up} & \Psi^\dagger_{\bq \down} \end{pmatrix}\begin{pmatrix} H_{\up\up} & H_{\up\down} \\ H_{\down\up} & H_{\down\down}\end{pmatrix}\begin{pmatrix} \Psi_{\bq \up} \\ \Psi_{\bq \down} \end{pmatrix} . \label{eqapp:hamexpand}
\end{gather}
Our objective here is to obtain the Hamiltonian of the $\Psi_{\bq \down}$ field to lowest order in $\bq$. The $\Psi_{\bq \up}$ quasiparticles, i.e., the quasiparticles governed by $H_{\up\up}$, are high-energy quasiparticles since they have a pairing energy gap $\sim \Delta_0$. The coupling to the low-energy gapless quasiparticles $\Psi_{\bq \down}$ can be treated by doing perturbation theory in $1/\Delta_0$. The effective Hamiltonian $\widetilde{H}(\bq)$ of the low-energy gapless quasiparticles to lowest order in $(q_x,q_y)$ is then obtained from perturbation theory as (see for instance Ref. 104)
\begin{gather}
\widetilde{H}(\bq) = H_{\down\down} - H_{\down\up} (H_{\up\up})^{-1} H_{\up\down} + \ldots \label{eq:perturb}
\end{gather}

\subsection{$J=1$ superconductor with $C_6$ symmetry \label{ssec:J1C6}}

We start with a $C_6$-symmetric superconductor. The gap function $\Delta_\bk$ to linear $p$-wave order in spherical harmonics is given by Eq. (13) (see also Table II). The Hamiltonian $H_{\down\down} $ appearing in Eq. \eqref{eqapp:hamexpand} only contains the part coming from the normal state dispersion, and therefore, to capture the effect of pairing, we need the second term of Eq. \eqref{eq:perturb}.

The Hamiltonian $H_{\up\up}(\bq)$ is given by
\begin{eqnarray}
H_{\up\up}(\bq) &=&  \begin{pmatrix}   v_F q_z & 0 & 0& -2\Delta_0 \lambda_b  \\ 0 & - v_F q_z& 2\Delta_0\lambda_b &0 \\ 0& 2\Delta_0\lambda_b  &  v_F q_z & 0 \\  -2\Delta_0\lambda_b  & 0&0& - v_F q_z\end{pmatrix} \nonumber \\
&=&  v_F q_z \sigma_z + \tilde{\Delta}_0  \sigma_y\tau_y, \label{eq:Hupup}
\end{eqnarray}
where $  \tilde{\Delta}_0  = 2 \Delta_0 \lambda_b$ is the energy scale associated with the $\Psi_{\bq \up}$ particles. The Hamiltonian block $H_{\up\down}(\bq)$ of Eq. \eqref{eqapp:hamexpand} which describes the coupling of the low- and high-energy degrees of freedom takes the form
\begin{eqnarray}
H_{\up\down}(\bq) &=& H_{\down\up}(\bq)  =  \frac{ \Delta_0\lambda_a}{\kf} \begin{pmatrix}   0 & 0 & 0& q_+ \\ 0 &0 &q_+& 0\\ 0& q_- & 0  &0\\ q_-  & 0&0&0 \end{pmatrix} \nonumber \\
& = &  \frac{\Delta_0}{k_F}\lambda_a \sigma_x( q_x\tau_x - q_y \tau_y) .
\end{eqnarray}

Using Eq. \eqref{eq:perturb} and expanding $(H_{\up\up})^{-1}$ to lowest order in $\Delta^{-1}_0$, and keeping terms up to order $\mathcal{O}(q^2)$ from the second term in Eq. \eqref{eq:perturb} we find the result
\begin{gather}
\widetilde{H}(\bq) = v_F q_z \sigma_z - \frac{1}{2m_\Delta}\sigma_y[(q^2_x-q^2_y)\tau_y + 2q_xq_y\tau_x ], \label{eqapp:hamC6}
\end{gather}
where the effective mass $m_\Delta$ for the $\Psi_{\bq \down}$ quasiparticles defined in Eq. (11) is given by
\begin{gather}
\frac{1}{m_\Delta}= \frac{ \lambda^2_a\Delta_0}{ k^2_F\lambda_b}.
\end{gather}

\subsection{$J=1$ superconductor with $C_3$ symmetry \label{ssec:J1C3}}

We proceed to consider the $C_3$-symmetric chiral superconductor. The gap function to leading $p$-wave order is given by Eq. (16), which differs from Eq. (13) by the admixture of the gap function component $i k_- s_-$. The latter component belongs to the $J=-2$ mod $6$ channel of a $C_6$-symmetric chiral superconductor.

We simply find that the low-energy gapless quasiparticles $\Psi_{\bq \down} $ are governed by the Hamiltonian $H_{\down\down}(\bq)$ to lowest order in $\bq$. Using Eq. (16) we find $H_{\down\down}(\bq)$ to linear order in $\bq$ as
\begin{gather}
H_{\down\down}(\bq) = \begin{pmatrix}  v_F q_z & 0 & 0& v_\Delta iq_- \\ 0 &-v_F q_z& v_\Delta iq_-& 0 \\ 0& -v_\Delta iq_+ & v_F q_z &0 \\  -v_\Delta iq_+& 0&0& -v_F q_z\end{pmatrix} \nonumber \\
= v_F q_z \sigma_z + v_\Delta \sigma_x ( q_y\tau_x- q_x\tau_y ). \label{eq:}
\end{gather}
The effective Fermi velocity $v_\Delta$, which is a property that characterizes the low-energy Majorana quasiparticles and was defined in Eq. (10), is expressed in terms of the gap function parameters as
\begin{gather}
v_\Delta \equiv 2\Delta_0\lambda_c/k_F.
\end{gather}

\section{Two-gap feature in DOS \label{app:DOS}}

In this section we demonstrate the two-gap feature in density of states (DOS). The energy spectrum in non-unitary superconductors is non-degenerate, thus leading to a distinctive two-gap feature in the total density of states. The easiest way to see it is to consider Hamiltonian (35) and put $\lambda_a = 0$ in the pairing potential of Eq. (16). In this case spin up and spin down sectors are completely decoupled. For spin up, the pairing potential has the form of polar phase of $^3\text{He}$, $\Delta_{\uparrow}=2\Delta_0\lambda_bk_z\tau_x/\kf,$ while spin-down sector possesses gap structure of $^3\text{He}$ $A$-phase, $\Delta_{\downarrow}=2\Delta_0\lambda_c (-k_x \tau_y+k_y\tau_x)/\kf$. The corresponding densities of states are equal to
\begin{align}
&\rho_1(\ve) = N_0 \frac{\ve}{2\Delta_c}\ln \frac{\ve+\Delta_c}{|\ve-\Delta_c|}, \qquad &A-\text{phase} \nonumber \\ &\rho_2(\ve) = N_0 \frac{\ve}{\Delta_b}\text{Re}\left( \arcsin\frac{\Delta_b}{\ve} \right), \qquad &\text{polar phase}
\end{align}
where we defined $\Delta_{b(c)} = 2\Delta_0|\lambda_{b(c)}|,$ $N_0 = m \kf/2\pi^2$ is the normal density of states per one spin projection. The total density of states $\rho(\ve) = \rho_1(\ve) + \rho_2(\ve) $, indeed, has two distinctive peaks at $\ve=\Delta_b$ and $\ve = \Delta_c$.

Analogously, in the other limiting case $\lambda_b = 0,$ $\lambda_{a,c} \ne 0,$ total density of states is given by $\rho(\ve) = \rho_1(\ve) + \rho_2(\ve) $, with
\begin{align}
&\rho_{1,2}(\ve) = N_0 \frac{\ve}{2\Delta_{1,2}}\ln \frac{\ve+\Delta_{1,2}}{|\ve-\Delta_{1,2}|}, \nonumber \\ &\Delta_{1,2}=\Delta_0(\sqrt{\lambda_a^2 + \lambda_c^2}\pm \lambda_c),
\end{align}
expressing the same two-gap behavior.

We emphasize that this two-gap feature appears in the single-band model and is purely due to non-unitary nature of pairing potential.

\section{NMR Calculations}

In this section, for completeness, we first briefly review the NMR relaxation rate calculation using Fermi's Golden Rule for unitary pairing and then consider non-unitary pairing corresponding to Eq. (13).

\subsection{Unitary pairing}

In case of unitary pairing, defined by $\Delta_{\bk} \Delta^\dagger_{\bk} \propto I_2$ [see Eq. (37)], the BdG Hamiltonian of Eq. (35) is diagonalized by a unitary transformation $\hat U$ which has the simple form
\be
\hat U = \begin{pmatrix} \hat u_{\bk} &  \hat v_{\bk} \\  \hat v_{-\bk}^* & \hat u_{-\bk}^*  \end{pmatrix},
\ee
with matrices $\hat u_{\bk}$ and $\hat v_{\bk}$ given by
\beq
\hat u_{\bk}  &=  &\frac{(E_{\bk} + \xi_{\bk} ) \cdot I_2}{\left[ (E_{\bk} + \xi_{\bk})^2 +  \Delta_{\bk}^2 \right]^{1/2}}  = u_\bk  \cdot I_2,  \nonumber \\
  \hat v_{\bk} & =&  \frac{- \Delta_{\bk}}{\left[ (E_{\bk} + \xi_{\bk})^2 +  \Delta_{\bk}^2 \right]^{1/2}} = - v_\bk \Delta_{\bk} . \label{eqapp:ukvk}
\eeq
Here we defined $\Delta_{\bk}^2 = (1/2) \Tr \Delta_{\bk}\Delta_{\bk}^\dagger$, and Bogoliubov quasiparticle energy spectrum is given by $E_{\bk} = \sqrt{\xi^2_{\bk} + \Delta_{\bk}^2}$. Using the matrix $\hat U$ which relates electron and hole operators to the Bogoliubov quasiparticle operators $a_{\bk }$, the hyperfine coupling Hamiltonian (21) can be reexpressed in terms of the Bogoliubov quasiparticle operators. In such an expression, all anomalous terms of the form $\sim a_\bk a_{\bk'} $ and $\sim a_{\bk}^\dagger a_{\bk'}^\dagger$ can be neglected as they do not contribute to the relaxation rate due to energy conservation (the delta-function in Fermi's Golden Rule).

The next step is to calculate the (square of the) matrix elements $\langle -\bS , a_{\bk s}   |\mathcal H_{\text{hf}}| \bS ,a_{\bk' s' }\rangle $. It follows from straightforward calculation that cross-terms, i.e., terms involving different components of nuclear spin $\hat \bS$, sum to zero due the unitary condition $\bd_{\bk} \times \bd_{\bk}^* =0$ [see Eq. (37) and subsequent discussion]. Then, using the expressions for the nuclear spin matrix elements given by $|\langle -\bS | \hat S_z | \bS \rangle|^2 = S_{\perp}^2/4$ and $|\langle -\bS | \hat S_{\pm} | \bS \rangle|^2 = (1\mp S_{z})^2/4,$ one finds for the relaxation rate
\begin{multline}
\frac1{T_1} = 2\pi(\gamma_NA_{\text{hf}})^2\sum_{\bk \bk'}f_{\bk'}(1-f_{\bk }) \delta(E_{\bk} -E_{\bk'}) \\
\times \left( |u_{\bk}|^2|u_{\bk'}|^2 +|v_{\bk}|^2|v_{\bk'}|^2 |\bd_{\bk}|^2|\bd_{\bk'}|^2\right), \label{eqapp:T1unitary}
\end{multline}
with $u_\bk, v_\bk$ defined in Eq. \eqref{eqapp:ukvk}.
Observe that this does not depend on $S_\perp$ and $S_z$. Making use of the identities $(E_{\bk} + \xi_{\bk})^2 + \Delta_{\bk}^2 = 2E_{\bk}(E_{\bk}+\xi_{\bk})$ and $ |\bd_{\bk}|^2 = \Delta_{\bk}^2 = E_{\bk}^2 - \xi_{\bk}^2$, we arrive at
\be
|u_{\bk}|^2|u_{\bk'}|^2 +|v_{\bk}|^2|v_{\bk'}|^2 |\bd_{\bk}|^2|\bd_{\bk'}|^2 = \frac12\left( 1+\frac{\xi_{\bk}\xi_{\bk'}}{E_{\bk}E_{\bk'} } \right).
\ee
Substituting this into \eqref{eqapp:T1unitary}, the second term vanishes after the momentum summations, and after standard manipulations one ends up with the final expression for the relaxation rate for unitary pairing given by
\be
\frac1{T_1}  = \pi (\gamma_N A_{\text{hf}})^2 \int dE f_{E}(1-f_{E})\rho^2(E).
\ee
Here the density of quasiparticle states $\rho(E)$ is defined as $\rho(E) = \sum_{\bk} \delta(E-E_{\bk}).$

The final expression shows that in case of unitary pairing $1/T_1$ does not have a dependence on the initial polarization of the nuclear spin. The reason can be traced back to the spin-degeneracy of the Bogoliubov quasiparticles spectrum.

\subsection{Non-unitary pairing}

The calculation of the NMR relaxation rate for non-unitary pairing states proceeds along the same lines. As an example, we consider $J=1$ chiral superconductor in crystals with $C_6$ symmetry and gap function (13) (without loss of generality we choose $\lambda_a,\lambda_b>0$).

The first step is to obtain the unitary transformation $\hat U$ which diagonalizes the BdG Hamiltonian. It takes the general form $\hat U = \{ {\bf u}_{1+}, {\bf u}_{2+}, {\bf u}_{2-},{\bf u}_{1-} \}$, where ${\bf u}_{1,2\pm}$ are (normalized) eigenvectors corresponding to the BdG energies $E_{\bk 1,2\pm}= \pm E_{\bk 1,2} $. The explicit expressions for the eigenvectors are rather involved, however, since we only need the low-energy part of the spectrum for our purposes (i.e., at low temperatures only the low-energy gapless excitations contribute to $1/T_1$), we can work with the simpler approximate expressions for the low-energy eigenstates. The branch of quasiparticle states with energy $E_{\bk 2}$ is gapped and we therefore neglect it altogether. Of the $E_{\bk 1}$ branch, we only keep the low-energy quasiparticle states near the nodes at $\pm \bK$ and expand the eigenvectors ${\bf u}_{1+}$ and ${\bf u}_{1-}$ in powers of small momentum $k_{\perp}$. Specifically, written explicitly in components, the initial electron operators can be expressed in terms of the Bogoliubov quasiparticles $a_{\bk1}$ as
\beq
c_{\bk \uparrow} &=& u_{\bk\up} a_{\bk1} - v_{\bk \up} a_{-\bk1}^\dagger , \nonumber \\
c_{\bk \downarrow} &=&u_{\bk\down} a_{\bk1} - v_{\bk\down} a_{-\bk1}^\dagger, \label{eqapp:transform}
\eeq
with the coefficients $u_{\uparrow(\downarrow)\bk}$ and $v_{\uparrow(\downarrow)\bk}$ given to the leading order by
\begin{align}
 & u_{ \bk\up} \approx  \frac{\lambda_a k_{\perp}}{2\lambda_b | k_z|}\sqrt{\frac12\left(1 + \frac{\xi_{\bk}}{E_{1\bk}}\right)},  \nonumber \\ & v_{ \bk\up} \approx  \frac{\lambda_a k_{\perp}}{2\lambda_b k_z}\sqrt{\frac12\left(1 - \frac{\xi_{\bk}}{E_{1\bk}}\right)}, \nonumber  \\ & u_{\bk\down } \approx  \frac{k_zk_+}{| k_z | k_{\perp}}\sqrt{\frac12\left(1 + \frac{\xi_{\bk}}{E_{1\bk}}\right)},  \nonumber \\ & v_{\bk\down} \approx  \frac{ k_+}{ k_{\perp} }\sqrt{\frac12\left(1 - \frac{\xi_{\bk}}{E_{1\bk}}\right)},
\end{align}
and the excitation energy equals $E_{1\bk} \approx (\xi^2_{\bk} + \lambda_a^4 k_{\perp}^4/4\lambda_b^2 k_z^2)^{1/2}$. Near the nodes, $k_z\approx \pm \kf$, from which we see that the coupling of spin-up electrons to Bogoliubov quasiparticles is $k_{\perp}/\kf \ll 1$ times weaker than the coupling of spin-down electrons. This is the origin of the strong anisotropy of NMR relaxation rate.

Using \eqref{eqapp:transform}, the hyperfine interaction Hamiltonian (21) can then be written in terms of the Bogoliubov quasiparticles as follows
\begin{multline}
\mathcal H_{\text{hf}} \cong \gamma_N A_{\text{hf}}  \sum_{\bk, \bk'} a_{\bk1}^\dagger a_{\bk'1} \left[ \hat S_+(u_{\da \bk}^* u_{\ua \bk'} - v_{\da -\bk'}^* v_{\ua -\bk}) \right. \\
+ \hat S_z(u_{\ua\bk}^*u_{\ua \bk'} - v_{\ua -\bk'}^* v_{\ua-\bk} - u_{\da\bk}^*u_{\da \bk'} + v_{\da -\bk'}^* v_{\da-\bk})  \\
\left. + \hat S_-(u_{\ua \bk}^* u_{\da \bk'} - v_{\ua -\bk'}^* v_{\da -\bk}) \right] , \label{eqapp:Hhf}
\end{multline}
where the anomalous terms $a_{\bk} a_{\bk'}$ and $a_{\bk}^\dagger a_{\bk'}^\dagger$ have been ignored. Substituting the expressions for $u_{ \bk \up,\down}$ and $v_{ \bk \up,\down}$, one obtains Eqs. (23) and (24).

We are now in a position to calculate $1/T_1$ from Eq. (38), assuming a spherical Fermi surface for simplicity. Under this assumption, it is clear that cross terms do not contribute to the relaxation rate as they vanish upon averaging over the directions on Fermi surface. Ultimately, using the explicit expressions for matrix elements $|\langle -\bS | \hat S_z | \bS \rangle|^2 = S_{\perp}^2/4$ and $|\langle -\bS | \hat S_{\pm} | \bS \rangle|^2 = (1\mp S_{z})^2/4$, we find (at low temperatures given by $T\ll \Delta_0\min\{\lambda_a,\lambda_b\}$)
\be
\frac1{T_1} = (\gamma_N A_{\text{hf}} )^2 \left[f_1(T) S_{\perp}^2 + f_2(T) S_{z}^2\right],
\ee
where $f_1(T)$ is given by
\begin{multline}
f_1(T) = \frac{\pi}2 \sum_{\bk \bk'} \left(|u_{\da\bk}|^2 |u_{\da\bk'}|^2 + |v_{\da\bk}|^2 |v_{\da\bk'}|^2  \right)  \\
 \times  f_{\bk' }(1-f_{\bk }) \delta(E_{1\bk} -E_{1\bk'})
=  \frac{\pi\lambda_b^2 \kf^2}{96 \lambda_a^4 \vf^2} T^3
\end{multline}
and $f_2(T)$ is given by
\begin{multline}
 f_2(T) =  2\pi\sum_{\bk \bk'} \left(|u_{\ua\bk}|^2 |u_{\da\bk'}|^2 + |v_{\ua\bk}|^2 |v_{\da\bk'}|^2  \right)  \\
 \times f_{\bk' }(1-f_{\bk }) \delta(E_{1\bk} -E_{1\bk'})
= \frac{9\zeta(3)\lambda_b \kf}{8\pi^2 \lambda_a^4 \vf^2} T^4.
\end{multline}
Rewritten in terms of the effective low-energy parameters $m_{\Delta} = \kf^2\lambda_b/\lambda_a^2\Delta_0 $ and $\tilde \Delta_0 = 2\Delta_0\lambda_b$, we obtain the final result quoted in Eq. (25).

\section{Semiclassical calculation of Majorana arc surface states}

In this section we present the calculation of the Majorana arc surface states of the $C_6$- and $C_3$-symmetric superconductors within the semiclassical approach. 

\subsection{$C_6$-symmetric superconductor}

We first consider the $J=1$ odd-parity chiral superconductor with hexagonal symmetry. In this case, the gap function is given by Eq. (13). We take $\lambda_a>0$ for definiteness; the case $\lambda_a<0$ is fully analogous.

For the gap function (13) the Hamiltonian $H_0(\bk_\parallel)$ takes the form
\begin{gather}
H_0(\bk_\parallel) = \frac{\Delta_0}{k_F}  [ \lambda_a k_x s_z\tau_x + \lambda_bk_z(s_x \tau_x - s_y \tau_y)], \label{eqapp:HparallelC6}
\end{gather}
and we find the Hamiltonian $H_\perp(k_\perp,-i\nabla_y) $ as
\begin{gather}
 H_\perp(k_\perp,-i\nabla_y)= k_\perp \left[ \tau^z \frac{-i\nabla_y}{m}   - \frac{\Delta_0}{k_F}\lambda_a s^z \tau^y  \right]. \label{eqapp:HperpC6}
\end{gather}
Solving $H_\perp(k_\perp,-i\nabla_y) \chi(y) \Psi_0=0$ is straightforward and one finds $\chi(y) = e^{-\kappa  y}$ with $\kappa  = \Delta_0  \lambda_a/\vf$. The corresponding spinor solutions $\Psi_{0,\alpha }$ ($\alpha=1,2$) are given by
\begin{gather}
\Psi_{0,1 }= \frac12\begin{pmatrix} -1 \\ 1 \\ 1 \\ 1 \end{pmatrix}, \quad \Psi_{0,2 } = \frac i2\begin{pmatrix} 1 \\ 1 \\ -1\\ 1 \end{pmatrix}.
\end{gather}
The wavefunction that satisfies proper boundary condition, $\left.\Psi(\br)\right\vert_{y=0}=0$, is given then by
\begin{gather}
\psi_{\bk_\parallel}(\br ) = \frac{1}{N} e^{i \bk_\parallel\cdot\br_\parallel} \sin(k_\perp y) \Theta(y) e^{-\kappa y},
\end{gather}
where $N=k_\perp/2\sqrt{\kappa(k_\perp^2+\kappa^2)}\approx  1 /2\sqrt{\kappa}$ is the normalization constant.

Using the solutions of Eq. \eqref{eqapp:HperpC6}, we return to a second-quantized formulation of the low-energy surface degrees of freedom by introducing the Majorana operators $\gamma_{\bk_\parallel 1,2} $ as
\begin{eqnarray}
\gamma_{\bk_\parallel 1}  &=& (c_{\bk_\parallel\up} + c^\dagger_{-\bk_\parallel\up}+c_{\bk_\parallel\down} +c^\dagger_{-\bk_\parallel\down})/2 \nonumber \\
\gamma_{\bk_\parallel 2} &=& i(c_{\bk_\parallel\up} + c^\dagger_{-\bk_\parallel\down}-c_{\bk_\parallel\down} - c^\dagger_{-\bk_\parallel\up})/2
\end{eqnarray}
It is clear that these operators satisfy the reality condition, i.e., $\gamma_{\bk_ \parallel \alpha}^\dagger = \gamma_{ -\bk_{\parallel} \alpha}$ ($\alpha = 1,2$), and indeed correspond to Majorana quasiparticles.

Finally, projecting Hamiltonian \eqref{eqapp:HparallelC6} into the subspace of Majorana operators $\gamma_{\bk_{\parallel}} = (\gamma_{\bk_{\parallel}1}, \gamma_{\bk_{\parallel}2})^T$ we obtain
\be
 \mathcal H_{\parallel} = -\frac{\Delta_0}{{2}\kf}\sum_{\bk_{\parallel}}  \gamma_{-\bk_{\parallel}}^T \left( \lambda_a k_x I_2 + \lambda_b k_z \tilde s_z  \right) \gamma_{\bk_{\parallel}}, \label{eqapp:HsurfaceC6}
\ee
where $\tilde s_z = \pm 1$ labels the surface Majorana degree of freedom. The zero-energy profile in the vicinity of $\bk_{\parallel}=0$ is given then by $|\lambda_a k_x | = |\lambda_b k_z|$.

\subsection{$C_3$-symmetric superconductor}

In the $C_3$ symmetric case the order parameter $\Delta_\bk$ is given by Eq. (16). In this case, the Hamiltonian $H_0(\bk_\parallel) $ is given by
\begin{multline}
H_0(\bk_\parallel) = \frac{\Delta_0}{k_F}  [ \lambda_a k_x s_z\tau_x + \lambda_bk_z(s_x \tau_x - s_y \tau_y) \\ + \lambda_c k_x (s_y\tau_x - s_x\tau_y)], \label{eqapp:HparallelC3}
\end{multline}
whereas the operator $H_\perp(k_\perp-i\nabla_y)$ takes the form
\begin{multline}
 H_\perp(k_\perp,-i\nabla_y)= k_\perp \left[ \tau^z \frac{-i\nabla_y}{m}   - \frac{\Delta_0}{k_F}\lambda_a s^z \tau^y  \right. \\ \left. + \lambda_c k_\perp(s_y \tau_y + s_x \tau_x)\right].
\end{multline}
Solving $H_\perp(k_\perp, -i\nabla_y)$ one can easily find the bound-state solutions as
\be
\Psi_{\alpha}(\br) = \frac{1}{N_{\alpha}} e^{i \bk_\parallel\cdot\br_\parallel} \sin(k_\perp y) \Theta(y) e^{-\eta_\alpha \kappa  y}\Psi_{0,\alpha}, \label{eqapp:surfacesolutionC3}
\ee
with dimensionless $\eta_{1,2} = [1 +  ( \lambda_c/ \lambda_a)^2]^{1/2} \pm \lambda_c/\lambda_a$ (recall $\kappa = \Delta_0\lambda_a/\vf$), and the spinors $\Psi_{0,\alpha}$ given by
\begin{gather}
\Psi_{0,1 }= e^{-i\pi/4}\begin{pmatrix} -1 \\ i \eta_1  \\ \eta_1 \\ i  \end{pmatrix}, \quad \Psi_{0,2 } = e^{i\pi/4}\begin{pmatrix} 1 \\  i \eta_2  \\ - \eta_2 \\  i  \end{pmatrix}.
\end{gather}
The normalization constants are equal to $N_{1,2} \approx \sqrt{\kappa}[1+ ( \lambda_c/ \lambda_a)^2]^{1/4}$.

Introducing a set of second-quantized Majorana operators $\gamma_{\bk_{\parallel}} = (\gamma_{\bk_{\parallel}1}, \gamma_{\bk_{\parallel}2})^T$ corresponding to the solutions \eqref{eqapp:surfacesolutionC3}, the effective Hamiltonian reads as $ \mathcal H_{\parallel} =(\Delta_0/ 2\kf )\sum_{\bk_{\parallel}} \gamma_{-\bk_{\parallel}}^T  H(\bk_\parallel)\gamma_{\bk_{\parallel}}$ with
\begin{gather}
H(\bk_\parallel) =  \frac{\lambda_c^2 - \lambda_a^2 }{\sqrt{\lambda_a^2+\lambda_c^2}}k_x I_2 + \lambda_c k_x \tilde s_z   + \frac{\lambda_a^2  \lambda_b k_z}{\lambda_a^2 +  \lambda_b^2 } \tilde s_x ,
\end{gather}
where again $\tilde s_z = \pm 1$ labels the surface Majorana degree of freedom. After the diagonalization we find
\begin{multline}
H(\bk_\parallel) =  \frac{k_x(\lambda_c^2-\lambda_a^2)}{\sqrt{\lambda_a^2+\lambda_c^2}}I_2  + \sqrt{\lambda_c^2k_x^2+ \frac{\lambda_a^4\lambda_b^2 k_z^2}{(\lambda_a^2+\lambda_c^2)^2}}\tilde s_z. \label{eqapp:surfaceHC3}
\end{multline}
The zero-energy profile in the vicinity of $\bk_{\parallel} = 0$ is given by $k_x^2(\lambda_a^2-3\lambda_c^2)(\lambda_a^2 + \lambda_c^2) = k_z^2\lambda_a^2\lambda_b^2$.
It follows from this equation that the zero-energy solutions at $\bk_{\parallel} = 0$ exist only provided $\lambda_a^2>3\lambda_c^2$. As a result, the Hamiltonian \eqref{eqapp:surfaceHC3} only has meaning under this condition.

\end{document}